\documentclass[useAMS,usenatbib]{mn2e}
\usepackage{color,graphicx} % NB you need to use graphicx or else you can't use include graphics command
\usepackage{latexsym, times, setspace, amssymb, amsfonts, rotating} 
\usepackage[fleqn]{amsmath}
\usepackage[integrals]{wasysym}
\usepackage[tight]{subfigure}
\usepackage{natbib}
\begin{document}

\title{On the Pulse Intensity Modulation of PSR~B0823$+$26}
\date{\today}

\author[N.~J.~Young et al.]{N.~J.~Young,$^1$\thanks{Email:
  young.neiljames@gmail.com} B.~W.~Stappers,$^1$ P.~Weltevrede,$^1$ 
  A.~G.~Lyne$^1$ and M.~Kramer$^{2,1}$\\ $^1$Jodrell Bank Centre for
  Astrophysics, The University of Manchester, Alan-Turing Building,
  Manchester M13 9PL, United Kingdom \\ $^2$Max-Planck-Institut
  f\"{u}r Radioastronomie, Auf dem H\"{u}gel 69, 53121 Bonn, Germany}
  \maketitle
\begin{abstract}
We investigate the radio emission behaviour of PSR~B0823$+$26, a
pulsar which is known to undergo pulse nulling, using an 153-d
intensive sequence of observations. The pulsar is found to exhibit
both short ($\sim$~min) and unusually long-term ($\sim$~hours or more)
nulls, which not only suggest that the source possesses a distribution
of nulling timescales, but that it may also provide a link between
conventional nulling pulsars and longer-term intermittent
pulsars. Despite seeing evidence for periodicities in the pulsar radio
emission, we are uncertain whether they are intrinsic to the source,
due to the influence of observation sampling on the periodicity
analysis performed. Remarkably, we find evidence to suggest that the
pulsar may undergo pre-ignition periods of `emission flickering', that
is rapid changes between radio-on (active) and -off (null) emission
states, before transitioning to a steady radio-emitting phase. We find
no direct evidence to indicate that the object exhibits any change in
spin-down rate between its radio-on and -off emission modes. We do,
however, place an upper limit on this variation to be $\lesssim6\,\%$
from simulations. This indicates that emission cessation in pulsars
does not necessarily lead to large changes in spin-down
rate. Moreover, we show that such changes in spin-down rate will not
be discernible in the majority of objects which exhibit short-term
($\lesssim1$~d) emission cessation. In light of this, we predict that
many pulsars could exhibit similar magnetospheric and emission
properties to PSR~B0823$+$26, but which have not yet been observed.

\end{abstract}

\section{Introduction}\label{sec:intro}
PSR~B0823$+$26 was one of the first pulsars to be discovered
\citep{cls68} and, as such, has been studied for over 40~yr. Despite
its typical spin parameters (i.e. rotational period $P\sim0.53$~s,
period-derivative $\dot{P}\sim1.71\times 10^{-15}$ and magnetic field
strength $B\sim0.96$~TG), this pulsar is by no means ordinary. Among
its most salient features, are its inter-pulse and post-cursor
emission components \citep{bbm73}, which are typically observed at a
few percent of its main-pulse peak intensity (\citealt{ran86}; see
also Fig.~\ref{0823limits}). These features are quite rare among the
pulsar population, particularly the inter-pulse emission which is only
observed among a small subset ($\sim3\,\%$; \citealt{mgr11}) of the
normal pulsar population. PSR~B0823$+$26 is also found to exhibit
pulsed, soft X-ray emission \citep{bwt+04}, which makes it only one
out of nine old pulsars ($1$~Myr~$<~\tau~<$~$20$~Myr) that produce
high-energy emission (e.g. \citealt{bwt+04,to05,lll08}).

In addition to these rare emission features, the object exhibits
abrupt cessation and re-activation of its radio emission
\citep{hw74,rit76,rr95}, that is commonly referred to as `pulse
nulling' \citep{bac70}. This phenomenon, which is also observed in
numerous other pulsars, affects all components of emission
\citep{bac70,rit76} and can be thought of as an extreme manifestation
of mode changing, that is a variation between active (radio-on,
hereafter) and quiescent (radio-off or null, hereafter) emission
modes. Through studying the emission behaviour of PSR~B0823$+$26 over
three observing runs ($\lesssim2$~h of observations in total),
\cite{rr95} infer a \emph{nulling fraction} (NF)~$-$~the fraction of
pulses which are radio quiet~$-$~that is $6.4~\pm~0.8\,\%$.

While the documented nulls in PSR~B0823$+$26 last a few pulse periods
\citep{rit76}, pulse nulling can be observed over a much wider range
of timescales in pulsars; that is, from just one or two pulses to many
days (e.g. \citealt{ran86,big92,klo+06,wmj07}). Accordingly, the NFs
of pulsars can also range from less than $1\,\%$ up to, in excess of,
$95\,\%$ \citep{dchr86, wmj07}. Due to this diversity in the observed
nulling statistics, and incomplete sample of known objects which
exhibit temporary radio emission failure~$-$~$\lesssim200$ nulling
(transient) pulsars are currently known, and even less have been
studied in detail~$-$~not much is known about the `typical' properties
of nulling pulsars (e.g. $P$, $\dot{P}$ or magnetic inclination angle
$\alpha$)\footnote{While the NF of a pulsar is found to be weakly
correlated with characteristic age \citep{ran86,wmj07}, no significant
correlation has been found with other basic pulsar properties or
geometry \citep{rw07}.}, nor why exactly they undergo such extreme
changes in their radio emission mechanism.

It has been suggested that `magnetospheric-state switching' could be
the underlying mechanism responsible for mode-changing and nulling in
pulsars (e.g. \citealt{bmsh82,con05,tim10}). This process refers to
alterations to the global current distribution in a neutron star
magnetosphere which, in turn, are thought to cause moding between
different emission states. These global, magnetospheric
reconfigurations are also proposed to explain the (quasi-)periodic
timing signatures in pulsar residuals through correlated changes in
spin-down rate ($\dot{\nu}$), such as those in the intermittent pulsar
B1931$+$24 \citep{klo+06,hlk10,lhk+10}. However, not much is known
about how such magnetospheric alterations are triggered, nor what
their physical timescales or periodicities should be; there are a
number of different models that attempt to explain this phenomenon,
e.g. non-radial oscillations \citep{rmt11}, asteroid belts
\citep{cs08}, precessional torques \citep{jon12}, surface temperature
variations in the polar gap region \citep{zqlh97} and magnetic field
instabilities \citep{grg03,ug04,rkg04,wmj07}, but none are conclusive.

With the above in mind, we present the analysis of recent high-cadence
observations of PSR~B0823$+$26, which were stimulated by the discovery
of longer than previously recorded nulls during regular timing
observations with the Lovell Telescope at Jodrell Bank.  These data
represent the most comprehensive study of the intermittent behaviour
of this pulsar for the first time. This subsequently enables the
comparison between long-term intermittent pulsars and more
`conventional' nulling pulsars, as well as insight into the
mechanism(s) which govern radio emission modulation in such
objects. In Section~\ref{sec:obs}, we describe the observations,
followed by an overview of the emission variability of the source in
Section~\ref{sec:mod}. In Section~\ref{sec:timing}, we review the
timing behaviour of the neutron star, and the simulation tool used to
model its rotational properties. Finally, we discuss the implications
of the results in Section~\ref{sec:discuss} and present our
conclusions in Section~\ref{sec:conc}.

%\vspace{-7mm}
\section{Observations}\label{sec:obs}
The observations of PSR~B0823$+$26, presented here, were obtained with
the Lovell Telescope over a period of approximately 153~d
(1~January~2009 to 3~June~2009). These data were obtained using the
Analogue~Filter~Bank (AFB) and Digital~Filter~Bank (DFB)\footnote{The
  DFB was commissioned approximately 20~d after the start of the AFB
  observations.}  back-ends using a 1400~MHz receiver, with an average
cadence of approximately 4 observations per day
(see~Table~\ref{tab:0823obs} for details).

In addition to this typical daily monitoring, we carried out three
observing runs with the AFB which spanned the entire time the source
is above the horizon at Jodrell Bank (see Table~\ref{tab:0823HCobs}
for details). These data were obtained to compare the long-term
variation with possible short timescale modulation. As a result, there
are significantly more observations with this back-end compared with
the DFB. Despite the deficit in the number of observations, the data
obtained with the DFB is complementary to the emission modulation
study due to the better sensitivity and wider bandwidth.

\begin{table}
\caption{System characteristics of the observations of PSR~B0823$+$26
described here. The bandwidth capability of the DFB was increased on
1~April~2009, i.e. $\mathrm{MJD}\sim54922.1$. Therefore, values quoted
for the DFB back-end are with respect to times prior to
(DFB$_{\mathrm{pre}}$) and post (DFB$_{\mathrm{post}}$) MJD~$\sim
54922.1$ respectively.}  
\centering
\begin{tabular}{ l l l l}
  \hline
  \hline
  System Property & AFB & DFB$_{\mathrm{pre}}$ & DFB$_{\mathrm{post}}$ \\
  \hline
  Time span of observations (d)          &  153.3 & 77.3   &  60.9     \\
  Total number of observations           &  1274  & 284    &  281      \\
  Typical observation duration (min)     &  6     & 6      &  6        \\
  Average observation cadence (d$^{-1}$) &  8.3   & 3.7    &  4.6      \\
  Typical sky frequency (MHz)            &  1402  & 1382   &  1374     \\
  Typical observing bandwidth (MHz)      &  32    & 113    &  128      \\
  Typical channel bandwidth (MHz)        &  1     & 0.25   &  0.25     \\
  \hline
\end{tabular}
\label{tab:0823obs}
\end{table}

\begin{table}
\caption{The observation properties for the three continuous observing
intervals during January~2009. Each continuous observing run has a
length $T_{\mathrm{obs}}$, for which there are $N_{\mathrm{obs}}$
separate observations.}  \centering
\begin{tabular}{ c c c }
  \hline
  \hline
  Start Epoch (MJD) & $T_{\mathrm{obs}}$~(d) &  $N_{\mathrm{obs}}$ \\ \hline 
  54837.7215        & 0.6390           &  152       \\
  54854.6844        & 0.6623           &  155       \\
  54857.7146        & 0.6067           &  145       \\
  \hline
\end{tabular}
\label{tab:0823HCobs}
\end{table}

%\vspace{-5mm}
\section{Emission Variability}\label{sec:mod}
\subsection{Nulling activity and flux limits}\label{sec:nulls}
The denser coverage provided by the AFB observations meant that they
were used for characterising the overall emission modulation
properties of PSR~B0823$+$26. To determine whether the pulsar was
radio-on or -off, we used the average profiles formed over the entire
bandwidth for each approximately 6-min long observation. This was done
by visual inspection and resulted in a time-series of one-bit data
corresponding to the `radio activity' of the pulsar i.e. 1's for
observations with detectable emission and 0's for observations with
non-detectable emission (i.e. radio-off states)\footnote{We note that
  only the nulling activity of the main-pulse component is considered
  here, as the 6-min observations do not provide enough sensitivity to
  characterise the inter-pulse or post-cursor emission
  properties.}. To complement the visual inspection, we also made use
of the timing model to confirm detections. A time-of-arrival (TOA) was
calculated for each observation and was compared to the known timing
model. Those in good agreement were confirmed as detections.

\begin{figure}
  \begin{center}
    % NB Trim dimensions are top, left, bottom, right
    \includegraphics[trim = 20mm 15mm 4mm 24mm, clip, height=8.5cm,width=5.5cm,angle=270]{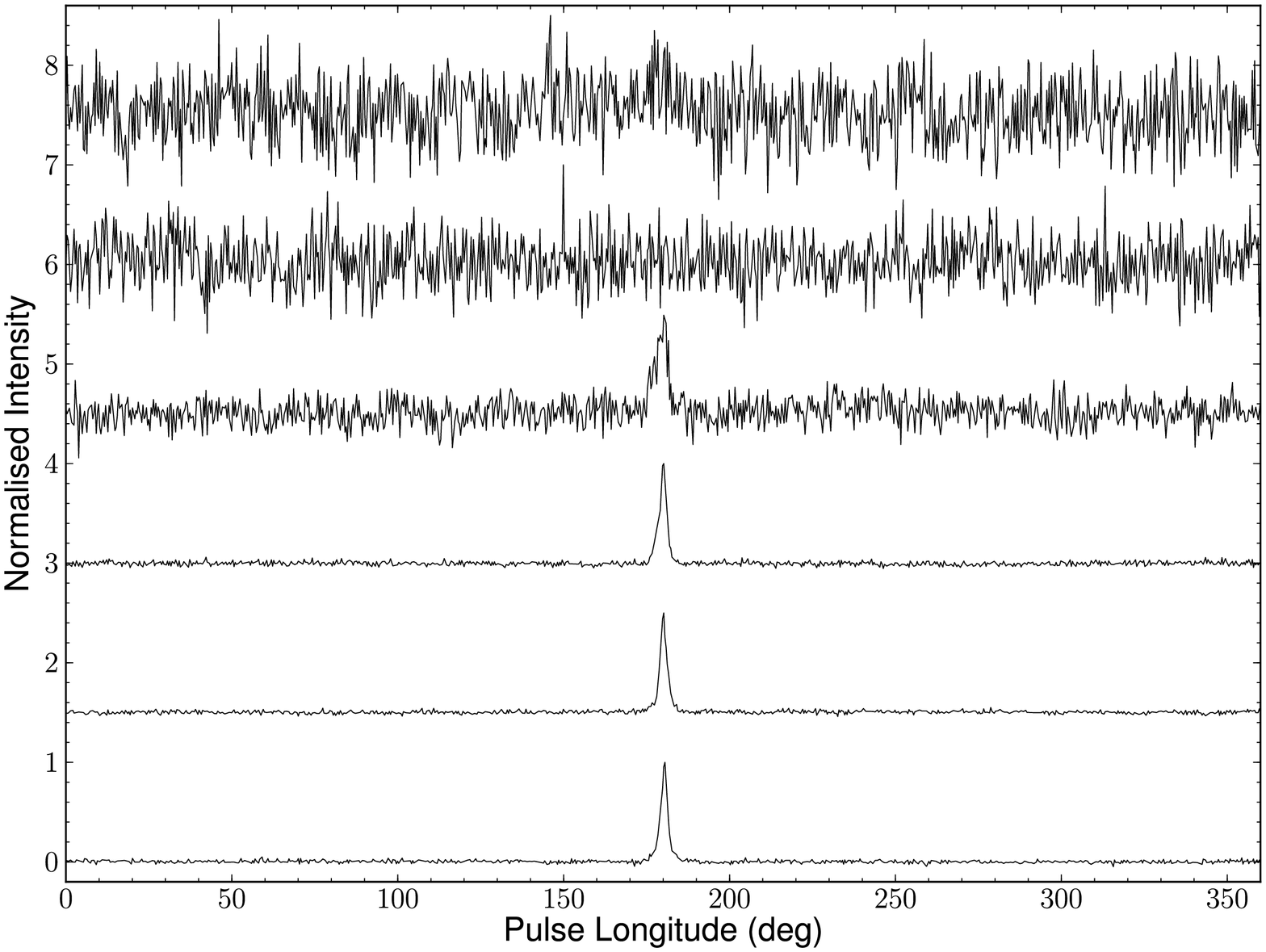}
  \end{center}
 \vspace{-4pt}
\caption{Consecutive pulse profile sub-integrations for PSR~B0823$+$26
  (\emph{from bottom to top}), which were obtained during one
  observation on 10~May~2009. The pulse intensity of each
  sub-integration is normalised to one and is offset from the next
  profile for clarity. The pulsar is detectable in the first four
  sub-integrations ($\sim60$~s each), after which it abruptly
  `switches off'. This transition occurs sometime during the fourth
  sub-integration over a timescale of less than one minute.}
\label{0823switch}
\end{figure}

Figure~\ref{0823switch} shows an example of an observation where the
pulsar is observed to transition between the radio-on and radio-off
phases. The transition timescales between the emission phases for this
pulsar are similar to those seen in PSR~B1931$+$24; that is, of the
order of seconds or less. Such sharp discontinuities in pulse
intensity are inconsistent with interstellar scintillation (see, e.g.,
\citealt{wmj+05} and references therein). Therefore, these transitions
between emission phases are considered to be intrinsic to the pulsar.

\begin{figure}
  \begin{center}
    % NB Normal trim dimensions are left,bottom,right,top 
    % NB Trim dimensions _here_ are bottom,right,top,left
    \includegraphics[trim = 3mm 10mm 2mm 3mm, clip,height=8.3cm,width=5.5cm,angle=90]{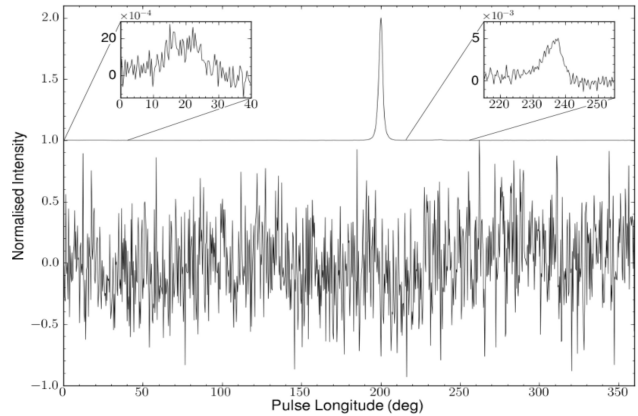}
  \end{center}
 \vspace{-7pt}
\caption{Average profiles of PSR~B0823$+$26 for the radio-on
  (\emph{top}) and radio-off (\emph{bottom}) observations
  respectively, which are offset for clarity. The maximum pulse
  intensities are normalised to one. The inset plots show zoom-ins of
  the inter-pulse (\emph{left}) and post-cursor (\emph{right})
  emission in the radio-on profile, using the same axes units as the
  main plot. Note that the y-axes of the inset plots represent the
  intensity offset from one.}
\label{0823limits}
\end{figure}

Due to the greater sensitivity, we use the DFB data to provide limits
on the average pulse flux density during the separate phases of
emission. By considering an integrated profile, formed from an
observation of length $T$, with an equivalent pulse width
$W_{\mathrm{eq}}$ and signal-to-noise ratio $\mathrm{SNR}$, the mean
flux density can be estimated via the modified radiometer equation
(e.g. \citealt{lk05}):
\begin{equation}
S = \frac{\beta \,\textrm{SNR} \,T_{\mathrm{sys}}}{G\sqrt{n_{\mathrm{p}}\,B\,T}}\sqrt{\frac{W_{\mathrm{eq}}}{P-W_{\mathrm{eq}}}}\,.
\label{eq:flux}
\end{equation}
Here, $\beta\sim1$ is the digitisation factor, $G\sim1$~JyK$^{-1}$ is
the telescope gain, $T_{\mathrm{sys}}\sim35$~K is the system
temperature, $n_{\mathrm{p}}=2$ is the number of polarisations,
$B=128$~MHz is the observing bandwidth, $P=531$~ms is the pulsar
period and \hbox{$W_{\mathrm{eq}}=10$~ms}. We averaged a total of
202~radio-on and 6~radio-off observations, which correspond to total
integration times of $T=1174.2$~min and $33.6$~min for the radio-on
and -off phases respectively. The resultant time-averaged profiles are
shown in Fig.~\ref{0823limits}. We place a limit on the mean flux
density in a radio-off phase $S_{\mathrm{off}}\leq0.022\pm0.004$~mJy
($\mathrm{SNR}\sim3$), which is approximately 100~times fainter than
that of the radio-on phase $S_{\mathrm{on}}=2.2\pm0.4$~mJy
($\mathrm{SNR}\sim1900$)\footnote{We note that the value quoted here
  for the radio-on flux density is lower than that obtained by
  \cite{lylg95} at 1400~MHz, i.e. $S_{1400}=10\pm2$~mJy. We attribute
  this discrepancy to our longer data set, which provides a more
  robust estimate due to the strong scintillation behaviour of the
  source (see, e.g., \citealt{wmj+05}).}. Converting these parameters
into pseudo-luminosities, using \citep{lk05}
\begin{equation}
L_{1400}\equiv S_{1400}\,d^2\,,
\label{eq:pseudolum}
\end{equation}
we find $L_{1400,\,\mathrm{off}}\lesssim2.9~\mu$Jy~kpc$^2$ and
$L_{1400,\,\mathrm{on}}\sim0.29$~mJy~kpc$^2$ (where the pulsar
distance $d\sim0.36$~kpc; \citealt{gtwr86}). We note that the
pseudo-luminosity of PSR~B0823$+$26 in the radio-off phase is at least
six times fainter than the weakest known radio pulsar PSR~J2144$-$3933
($L_{1400}\sim20~\mu$Jy~kpc$^2$; \citealt{lor94}). Although this
implies that the radio-off phases are consistent with emission
cessation, we cannot rule out the possibility that they may exhibit
extremely faint emission (c.f. \citealt{elg+05}).

\begin{figure*}
  \begin{center}
    % NB Trim dimensions are top, left, bottom, right
    \includegraphics[trim= 20mm 20mm 2mm 28mm,clip,angle=270,totalheight=15in,height=10.2cm,width=17.5cm]{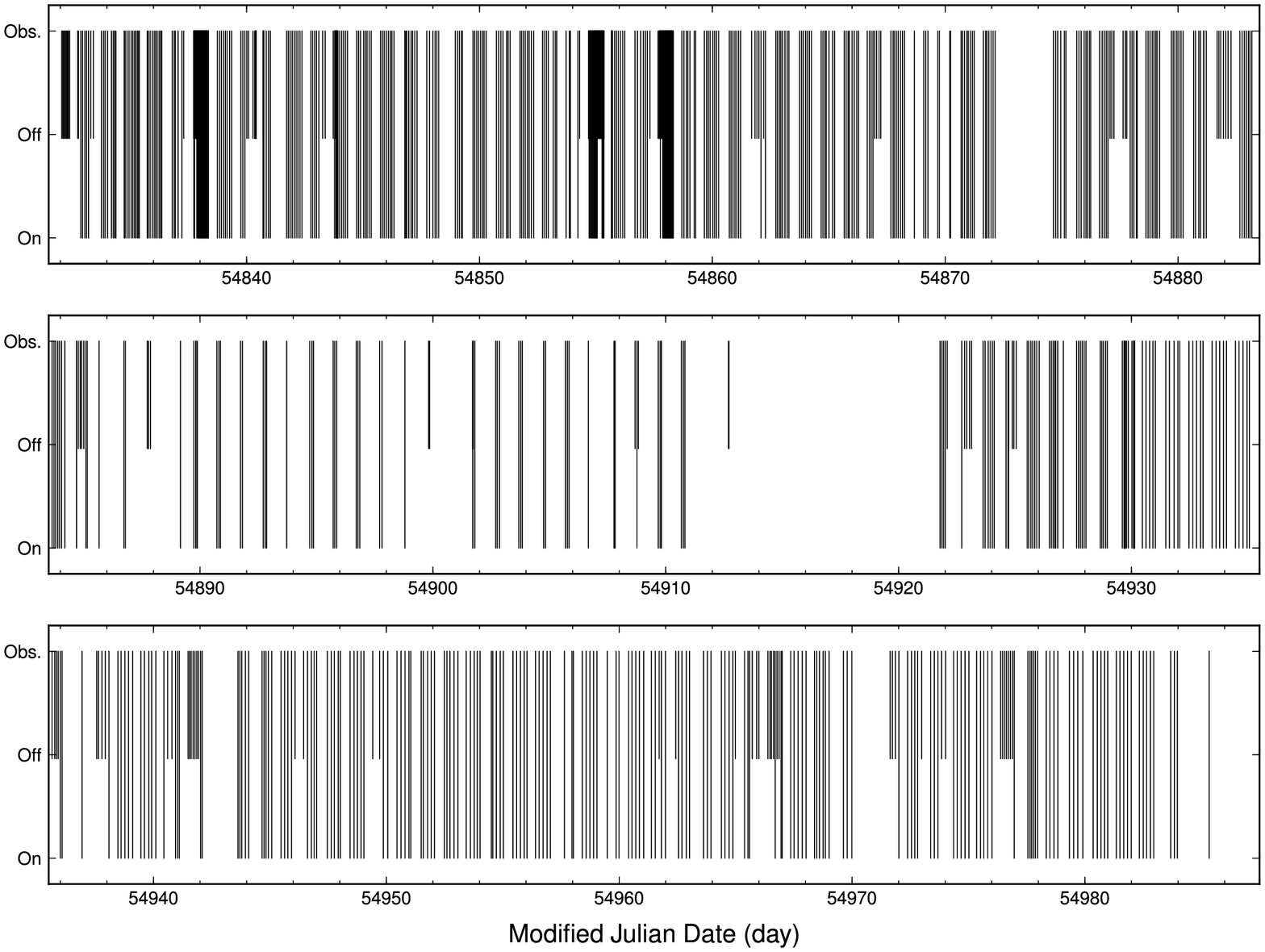}
   \end{center}
 \vspace{-5pt}
\caption{The sequence of observations of PSR~B0823$+$26 carried out
  over the 153-d period, denoted by the black lines. The data are
  separated into three continuous $N\sim54$~d panels. The times of
  observation and the times when PSR~B0823$+$26 was radio-on
  (full-amplitude) and -off (half-amplitude) are shown by the extent
  of the black lines. The times of more intensive observing sessions
  are shown at $\mathrm{MJD}\sim54838,54855$~and~$54858$.}
\label{0823onoff}
\end{figure*}

In Fig.~\ref{0823onoff}, we show an `activity plot', formed from the
153-d one-bit time-series data, which indicates when the pulsar was
observed and whether it was radio-on or -off. The radio emission in
PSR~B0823$+$26 is clearly observed to undergo modulation over variable
timescales, which are much shorter than those seen for PSR~B1931$+$24
(i.e. days to weeks; \citealt{klo+06}).

\begin{figure*}
  \begin{center}
    % NB Trim dimensions are top, left, bottom, right
    \includegraphics[trim= 20mm 19mm 2mm 32mm,clip,angle=270,totalheight=15in,height=10.2cm,width=18cm]{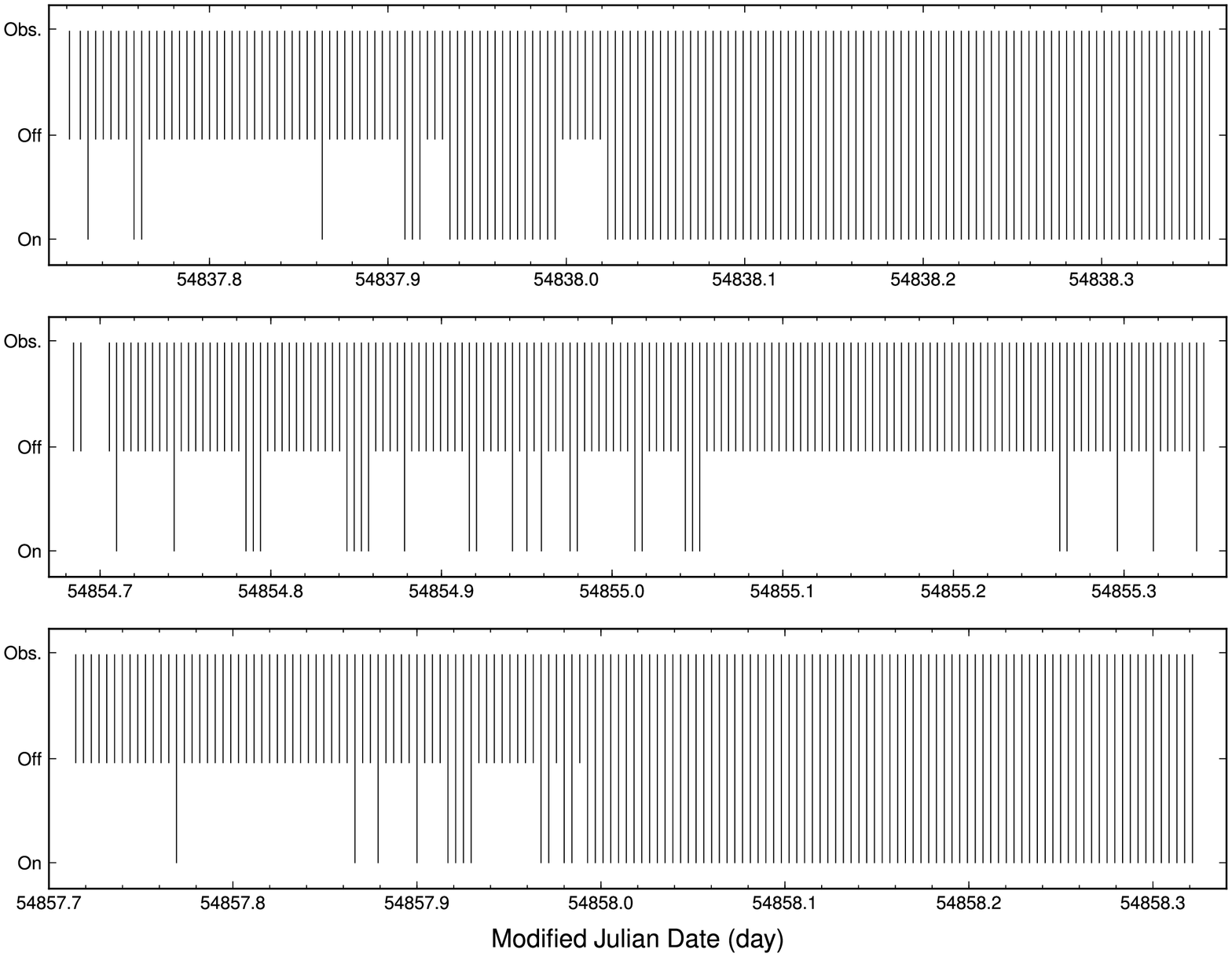}
   \end{center}
 \vspace{-5pt}
\caption{Three sequences of continuous observations of PSR~B0823$+$26
  during January~2009. The times when the pulsar was radio-on and -off
  are shown by the extent of the black lines.}
\label{0823shortscale}
\end{figure*}

The short-term emission modulation of PSR~B0823$+$26 was probed using
the data intervals of continuous observations, which are detailed in
Table~\ref{tab:0823HCobs}. These three data sets, shown in
Fig.~\ref{0823shortscale}, display some evidence for `emission
flickering'; that is, rapid changes between the radio-on and -off
states of emission before a constant emission mode is assumed. We
note, however, that further high-cadence observations are required to
confirm this, and the typical characteristics of the `flicker' pulses.

\begin{figure}
  \begin{center}
    % NB Trim dimensions are left,bottom,right,top 
    \includegraphics[trim = 10mm 20mm 14mm 28mm, clip,height=10cm,width=8.5cm]{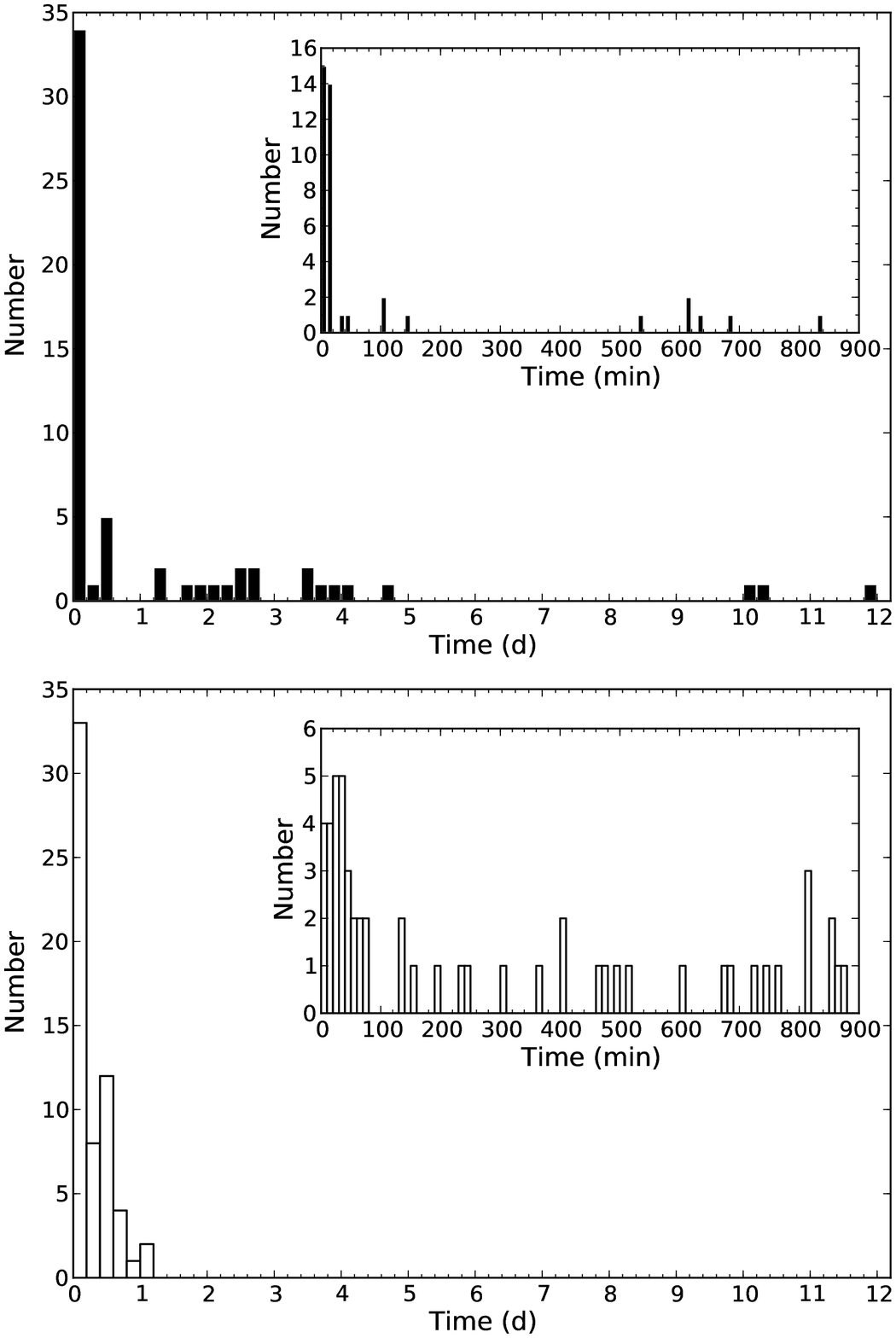}
    % NB Trim dimensions are top, left, bottom, right
%    \includegraphics[trim = 10mm 0mm 10mm 0mm, clip, height=14cm,width=6cm,angle=270]{figs/0823acthists_lscape.ps}
  \end{center}
 \vspace{-5pt}
\caption{Histograms showing the durations of time when the pulsar is
  observed to be in a specific emission phase, on (\emph{top}) and
  off (\emph{bottom}), with inset plots of the radio emission activity
  for timescales up to 15~h.}
\label{0823_acthists}
\end{figure}

We also determined the emission phase durations for the object, which
are defined as the difference between the start and end points
(i.e. transition times) of emission phases. To reduce systematic
error, the transition times for emission phases are assumed to be the
mid-points between consecutive radio-on and -off observations, that is
when the pulsar changes from radio-on to -off and vice
versa. Fig.~\ref{0823_acthists} shows the result of this
analysis. These data clearly show that the distribution of radio-on
phase durations exhibits a broader spread of values compared with that
of the radio-off phase. Overall, however, there is a bias towards
shorter emission phase durations, which is in stark contrast to
PSR~B1931$+$24. The average time that PSR~B0823$+$26 exhibited
detectable radio emission was $1.4\pm0.4$~d. Whereas, the average
radio-off timescale was $0.26\pm0.04$~d. The activity duty cycle
(ADC), the percentage of time in the radio-on phase, was calculated
from the ratio of the total radio-on duration to the total observation
time. The uncertainty in this value was determined from the ratio of
the standard error in the radio-off time to the mean radio-off
time. Subsequently, the pulsar is found to be radio-on for
$80\pm10~\%$ of the time.

We were also interested in determining whether the pulsar exhibits any
systematic trend in pulse intensity before (or after) a null
(c.f. PSR~B0809$+$74; \citealt{la83,vkr+02}). Ideally, this analysis
should be performed on single-pulse data. As these data were not
available, however, we were limited to using the 6-min integrated
pulse profiles to discern any systematic brightness variations in the
emission phases. We computed the peak flux density for each profile,
using the first term in Eqn.~\ref{eq:flux}, assuming a constant
observing system (i.e. with stable $T_{\mathrm{sys}}$, $\beta$ and $G$
parameters). We find no evidence to suggest any correlation between
the pulse intensity and pulse integration number preceding (or
following) a radio-off phase in these data. In addition, no
correlation was found between emission phase length and pulse
intensity.

%\vspace{-4mm}
\subsection{Periodicity analysis}\label{sec:pdcty}
In order to elucidate the behaviour of PSR~B0823$+$26, we have
performed \emph{weighted wavelet Z-statistic} (WWZ) analysis
\citep{fos96} on the one-bit time-series data. The WWZ algorithm
adopts a modified approach to traditional wavelet analysis (see, e.g.,
\citealt{add02} for a review of wavelet transforms) in order to
counter the undesired effects of uneven time sampling (e.g. spectral
leakage; \citealt{sca82}). It employs a wavelet function which
includes a periodic, sinusoidal test function, of the form
$e^{i\omega(t-\tau)}$, by projecting the data onto a set of sine and
cosine trial functions (i.e. the waveform). It also utilises a
Gaussian window function (weighting function of the data) which is
defined as \citep{fos96}
\begin{equation}
  w(\omega,\tau) = e^{-c\,\omega^2(t-\tau)^2} \,,
\label{eq:weights}
\end{equation}
and is centred at time $\tau$, with a width defined by the frequency
$\omega$ and tuning constant $c$.

As such, the WWZ uses the sinusoidal wavelet to fit the data and the
sliding window function to weight the data points which, in turn,
mitigates spectral leakage. We note here that data points towards the
centre of the window in the fit are weighted the heaviest, and those
near the edges of the window the least. The spectral content of a
signal is, subsequently, obtained at times corresponding to the centre
of the wavelet windows \citep{bzjf98,tem04,tmw05}.

\begin{figure*}
  \begin{center}
    %% Normal trim dimensions are left,bottom,right,top
    % Trim dimensions _here_ are top,left,bottom,right
    \includegraphics[trim = 15mm 0mm 10mm 0mm,clip,height=12.5cm,width=9cm,angle=270]{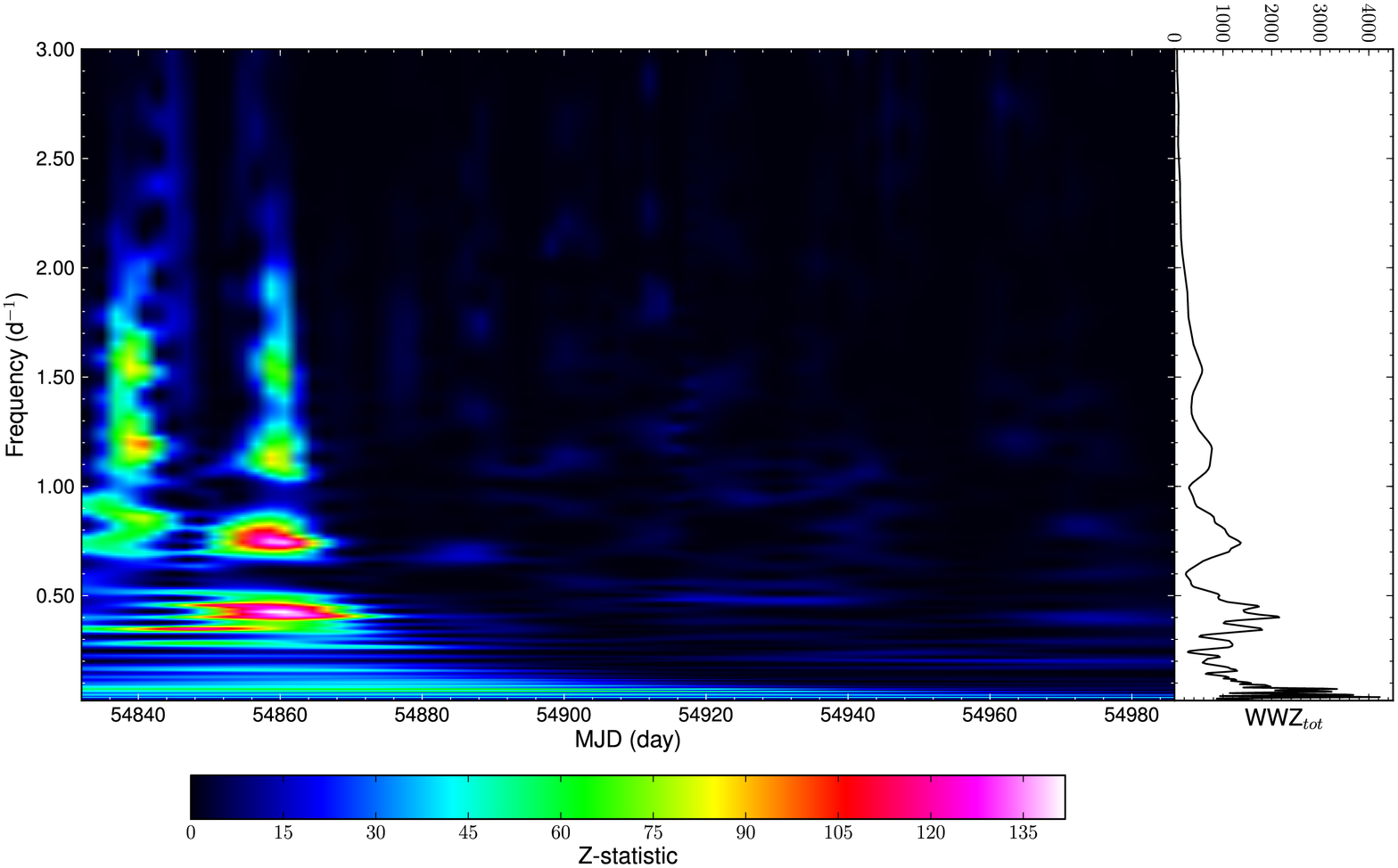}
  \end{center}
\vspace{-7pt}
\caption{WWZ transform of the 153-d PSR~B0823$+$26 radio emission
  activity data-set (\emph{left}) and the corresponding integrated
  power spectrum (\emph{right}). During the initial $\sim40$~d of the
  data-set, the observation sampling is sufficient to resolve several
  prominent features. These spectral components are also accompanied
  by a broad distribution of power towards higher frequencies, which
  are likely analogous to Fourier harmonics. At later times
  (\mbox{MJD~$\gtrsim54872$}), the reduced data sampling does not
  allow insight into the dominant fluctuation frequencies.}
\label{0823wwz}
\end{figure*}

For the purpose of our analysis, we chose $c=0.001$ so that we could
strike a balance between frequency and time resolution. The resultant
WWZ transform of the one-bit time-series data, showing the spectral
power at successive epochs (or \emph{time lags}), is displayed in
Fig.~\ref{0823wwz}. It is clear that the WWZ transform exhibits very
sporadic structure. This is thought to result from the effect of
irregular data sampling on the projection (i.e. local matching) of the
WWZ wavelet function. During the first $\sim40$~d of the data-set, the
observation sampling is at its greatest, with no gaps and $\sim5-10$
observations per day (or more). The three continuous observing
intervals are also included in this date range. Consequently, the data
obtained over the first $\sim40$~d accounts for $\sim62\,\%$ of the
total. We find that the data sampling in this observing period is
sufficient for the matched wavelet function to return power at several
fluctuation frequencies, which can be attributed to the locations of
the continuous observing sessions. This is highlighted by the split in
the WWZ data, centred around $\mathrm{MJD}\sim54948$, which shows that
the locations of dominant spectral features strongly correlate with
the midpoints of the continuous observing sessions. The most prominent
spectral feature can be seen at approximately $0.44-0.36$~d$^{-1}$,
corresponding to a period of about $2-3$~d. As the data sampling
worsens with time, becoming more irregular (roughly a few observations
per day) with occasional gaps (one is more than 10~d), there becomes a
point at which the prominent fluctuation frequencies of the system are
no longer resolved (i.e. around $\mathrm{MJD}\sim54972$).

In order to clarify the significance of these variations, we simulated
several data-sets using the observed data sampling. We created model
data-sets with single periodicities, ranging from 0.25~d to 10~d, and
analysed them with the WWZ. We also performed WWZ analysis on data
with randomised activity values. For the random data, the spectral
power is distributed across the entire WWZ plane and no dominant
periodicities are recovered. For the data with an intrinsic
periodicity, however, we find that the fundamental frequency is
resolved at virtually all epochs\footnote{Around
$\mathrm{MJD}\approx54913-54922$ there is a gap in the observations,
which often results in a prominent decrease in the WWZ
power.}. However, the fundamental spectral component is also
accompanied by a broad distribution of power and is substantially less
significant after the first 40~d\footnote{The power relating to a
time-frequency component in the WWZ represents its significance
\citep{bzjf98}.}. For simulated periods of the order of a day, we note
that the resulting distribution of power is similar to the broad
distribution observed in the real data. This indicates that our
results are modulated by data-windowing effects. We note that the
higher period components in the observed data ($\sim10-30$~d) are most
probably not intrinsic to the pulsar. This is because the analysing
wavelet has a full window-width ($\sim10$~cycles) that is of the order
of the length of the data-set and, therefore, is not suitable to
provide meaningful results about any long-term periodicity
\citep{tem04}. These results suggest that there are a number of
fluctuation frequencies present in the data, which are only resolved
during epochs when there is the most frequent observation
sampling. However, we are cautious to attribute these periodicities to
the intrinsic variability of the source, due to the influence of the
observation sampling on the results.

In order to elucidate the periodicities in the pulsar radio emission,
we now consider the error estimation of the WWZ data
(Fig.~\ref{0823wwz}). We note, however, that such error estimation is
a non-trivial procedure. This is highlighted by \cite{fos96}, who
state that analytic description of errors in a WWZ is very intricate,
due to the nature of the weighted parametric projection. In addition,
the assumptions made whilst devising the WWZ statistic practically
invalidate the formal errors. For example, we assume the null
hypothesis that the data is purely sinusoidal with constant frequency
and amplitude, plus random noise, which we know is false. Therefore,
any analytic errors which are calculated are consequently subject to
these assumptions. As a result, we have applied a more heuristic
approach to the estimation of WWZ errors. Here we employ two different
methods, both of which are complementary.

%\vspace{-4mm}
\subsubsection{Confusion limit estimation method}\label{sec:HWHM}
\begin{figure} 
  \begin{center}
    %% Normal trim dimensions are left,bottom,right,top
    % Trim dimensions _here_ are top,left,bottom,right
    \includegraphics[trim = 9mm 2mm 4mm 0mm,clip,height=8.5cm,width=5cm,angle=270]{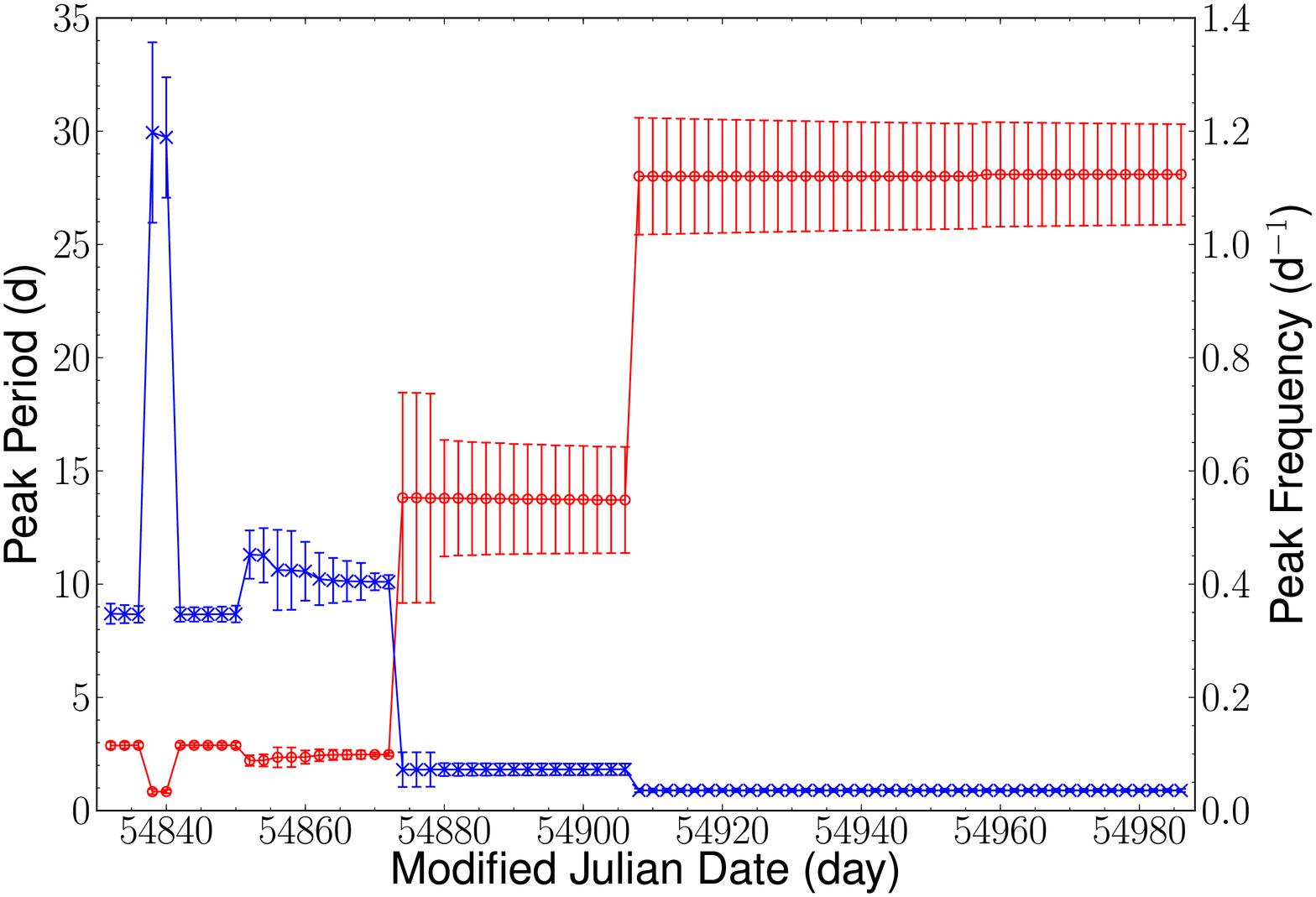}
  \end{center}
\vspace{-5pt}
\caption{The peak fluctuation frequencies (\emph{crosses}) and periods
  (\emph{open circles}) from the weighted wavelet Z-transform
  data. Error bars are $1$-$\sigma$ values computed using the
  confusion limit estimation method (fractional errors for individual
  epochs). The average error in the peak fluctuation frequency
  (period) is $\sim11\,\%$. For the first 40~d, the peak fluctuation
  period is typically $\sim2-3$~d apart from at a couple of epochs
  ($\sim0.8$~d), which may represent a variation in the intrinsic
  periodicity. After $\mathrm{MJD}\gtrsim54872$, the data sampling
  becomes poorer, and more irregular, which favours the smaller
  (non-physical) peak fluctuation frequencies.}
\label{0823periods}
\end{figure}

Following \cite{tmw05}, we used the \emph{confusion limit} method to
estimate the peak fluctuation frequencies and their \emph{maximum}
$1$-$\sigma$ uncertainties. Here, the peak fluctuation frequency at a
given epoch is one which is associated with the greatest WWZ
power. The maximum $1$-$\sigma$ uncertainty associated with this
frequency is approximated by measuring the confusion limit of the WWZ
spectrum, that is the half-width at half-maximum of the Z-statistic,
$Z(\omega,\,\tau)$. The result of this analysis is shown in
Fig.~\ref{0823periods}. We find that the average error in the
fluctuation frequency from this analysis is $\sim11\,\%$.

The variation in the dominant periodicity, within the first 40~d, is
above the error limit. However, as the data is clearly modulated by
data-windowing effects (see above), we believe that these variations
in the dominant periodicity are most likely governed by the
observation sampling, rather than any modulation in the pulsar radio
emission. Through comparing the observed data with the random
simulated data, we conclude that the source does exhibit some
(quasi-)periodicity, but which cannot be accurately resolved here due
to the aforementioned data-windowing effects.

%\vspace{-4mm}
\subsubsection{Data-windowing method}\label{sec:wwzwindow}
Uncertainties in the WWZ transform were also estimated by computing
the standard deviation of windowed data. Here, we separated the WWZ
data into segments of length $T=14$~d, and calculated the standard
deviation in the peak frequencies ($\sigma_{\nu(\tau)}$) and periods
($\sigma_{P(\tau)}$) for each segment accordingly. This provided a
measure of the modulation in the peak frequency and period over
time. We note, however, that we only included peak fluctuation periods
which were below 10~d in this analysis, as these periods are the
longest which are well represented by the WWZ (see above).

\begin{table*}
\caption{Summary of the results from the data-windowing error
  analysis. The time span of the WWZ data analysed is denoted by `MJD
  range'. The median-peak WWZ values, of these segements, and their
  $1$-$\sigma$ uncertainties are denoted by
  $\mathrm{WWZ}_{\mathrm{max}}$ and
  $\Delta(\mathrm{WWZ}_{\mathrm{max}})$ respectively. The peak
  fluctuation frequencies, periods and their corresponding
  $1$-$\sigma$ uncertainities are given by $\nu_{\mathrm{fluc}}$,
  $\Delta(\nu_{\mathrm{fluc}})$, $P_{\mathrm{fluc}}$ and
  $\Delta(P_{\mathrm{fluc}})$ respectively.}  \centering
\vspace{4pt}
\begin{tabular}{c  c  c  c  c  c  c }
  \hline
  \hline
   MJD range & $\mathrm{WWZ}_{\mathrm{max}}$ & $\Delta(\mathrm{WWZ}_{\mathrm{max}}$) &
   $\nu_{\mathrm{fluc}}$(d$^{-1}$) & $\Delta(\nu_{\mathrm{fluc}})$(d$^{-1}$) & $P_{\mathrm{fluc}}$(d) & $\Delta(P_{\mathrm{fluc}})$(d)\\
  \hline
54832$-$54846 &  94   &  25  &  0.35    &  0.39    &  2.9   &  1.0  \\
54846$-$54860 &  125  &  9   &  0.42    &  0.05    &  2.4   &  0.3  \\
54860$-$54874 &  113  &  32  &  0.40    &  0.12    &  2.5   &  4.0  \\
54874$-$54888 &  63.9 &  0.2 &  0.0728  &  0.0001  &  13.79 &  0.02 \\
54888$-$54902 &  62   &  1   &  0.0728  &  0.0001  &  13.75 &  0.02 \\
54902$-$54916 &  57   &  1   &  0.04    &  0.02    &  28.0  &  7.4  \\
54916$-$54930 &  57.4 &  0.3 &  0.0357  &  0.0000  &  28.0  &  0.0  \\
54930$-$54944 &  58.0 &  0.1 &  0.0357  &  0.0000  &  28.0  &  0.0  \\
54944$-$54958 &  58.1 &  0.1 &  0.03570 &  0.00004 &  28.01 &  0.03 \\
54958$-$54972 &  57.5 &  0.3 &  0.0356  &  0.0000  &  28.09 &  0.00 \\
54972$-$54986 &  56.3 &  0.5 &  0.0356  &  0.0000  &  28.09 &  0.00 \\
  \hline
\end{tabular}
\label{tab:0823tab}
\end{table*}

Table~\ref{tab:0823tab} shows the results of this analysis. The median
peak fluctuation frequencies (periods) for the first three segments of
the transform data are $\sim0.42-0.35$~d$^{-1}$ ($\sim2-3$~d). We
estimated the significance of periodicities within the WWZ data using
a bootstrapping approach (e.g. \citealt{st95,prw99,zi04}). Here, we
resampled the transform data 100 times and, for each resample,
determined the standard deviation. This allowed us to determine an
accurate estimate for the average standard deviation, or background
noise, of the transform data from the population of possible
values. We assume a $5\,\sigma_{\mathrm{WWZ}}\approx80$ level as a
confident signal detection. Consequently, we find that the median-peak
WWZ values of the windowed data are significant for
$\mathrm{MJD}\lesssim54872$. Whereas, those afterwards do not meet the
cut-off criteria. This further indicates the importance of high-time
resolution on the WWZ analysis; that is, periodicities in the data can
only be accurately constrained when sufficient time resolution is
available.

%\vspace{-4mm}
\subsubsection{WWZ analysis summary}\label{sec:wwzsum}
The above results indicate that PSR~B0823$+$26 may exhibit a number of
(quasi-)periodic features in its radio emission; the most prominent of
these being around $2-3$~d. However, evidence for their existence is
only obtained during a small portion of the data-set (i.e. in the
first 40~d), when the observation sampling is at its highest. From the
simulated data-sets, it also appears that these `significant' features
may be influenced by spectral leakage (i.e. data-windowing effects on
the WWZ). In light of this, we postulate that PSR~B0823$+$26 does
exhibit a fundamental (quasi)-periodicity in its radio emission, but
stress that further higher-cadence observations are required to better
characterise this behaviour.

\section{Timing Behaviour}\label{sec:timing}
\begin{figure}
  \begin{center}
    % Trim dimensions _here_ are top,left,bottom,right
    \includegraphics[trim = 10mm 7mm 4mm 15mm,clip,height=8.3cm,width=6cm,angle=270]{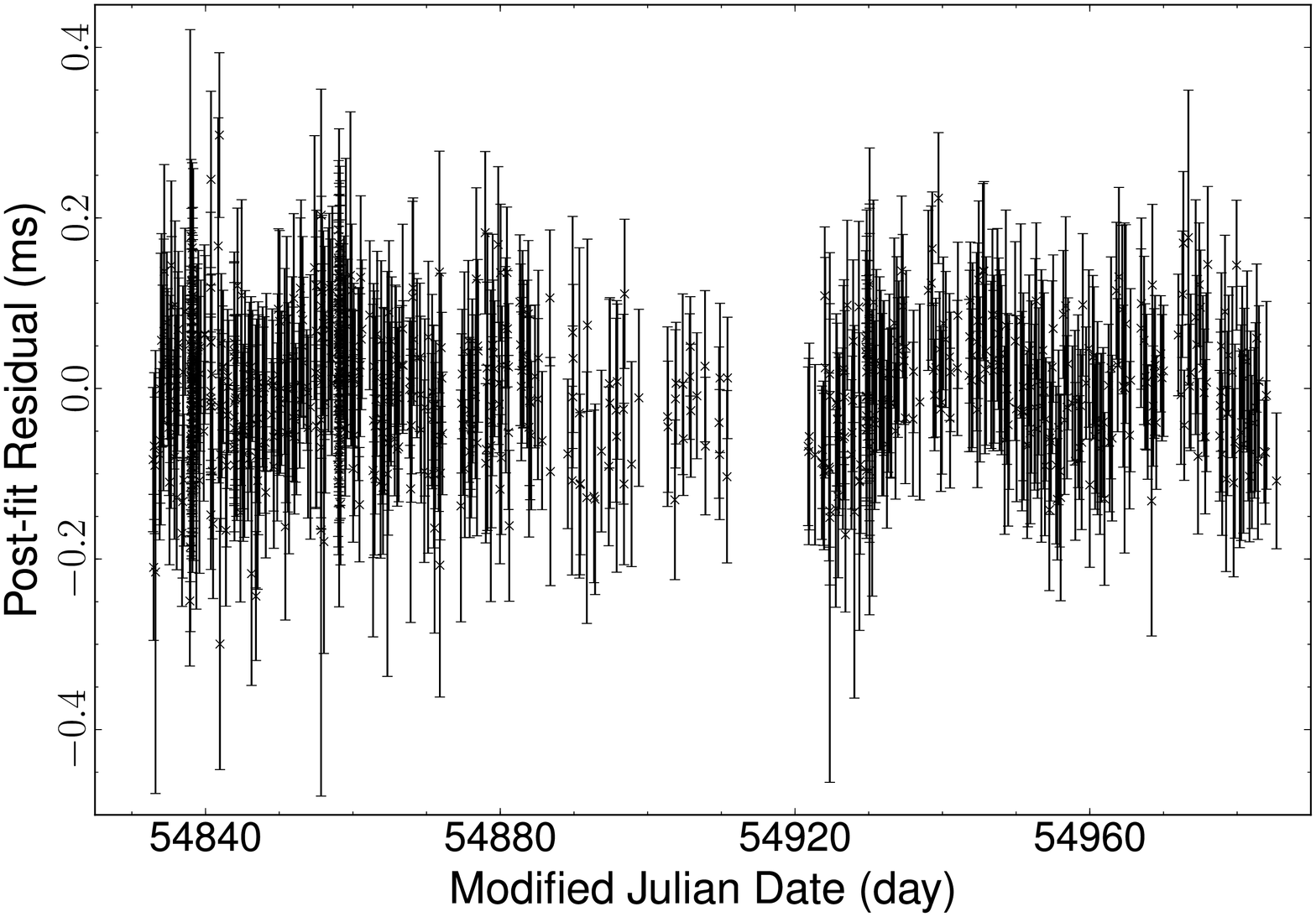}
  \end{center}
\vspace{-4pt}
\caption{Post-fit timing residuals for PSR~B0823$+$26 from the 153-day
  AFB data-set, after fitting for $\nu$ and $\dot{\nu}$. There is no
  apparent cubic structure in the timing residuals.}
\label{0823res}
\end{figure}

In PSR~B1931$+$24, we see systematic variations in its timing
residuals, which manifest as quasi-periodic cubic structure due to
spin-down rate variation \citep{klo+06}. To determine whether
PSR~B0823$+$26 exhibits similar behaviour, we analysed the timing
measurements obtained from the 153-d AFB data-set. We obtained the
topocentric TOAs using {\fontfamily{cmr}\footnotesize\selectfont
PSRPROF}\footnote{http://www.jb.man.ac.uk/~pulsar/observing/progs/psrprof.html}
and analysed them using the {\fontfamily{cmr}\footnotesize\selectfont
TEMPO2} package\footnote{A detailed overview of this timing package is
provided by \cite{hem06}. Further details and documentation can also
be found at http://www.atnf.csiro.au/research/pulsar/tempo2/.}. The
resulting residuals (observed~$-$~predicted TOAs) from a best-fit
timing model (Table~\ref{tab:0823par}) are shown in
Fig.~\ref{0823res}.

\begin{table}
\caption{The properties of PSR~B0823$+$26 obtained from timing
measurements of the 153-d AFB data-set. Note that the right ascension,
declination and dispersion measure of the source are held fixed in the
fit to these data. The standard $1$-$\sigma$ errors are provided in
the parentheses after the values, in units of the least significant
digit.}  
\centering
\begin{tabular}{ l l }
  \hline
  \hline
  Parameter                                    &  Value             \\ 
  \hline
  Right Ascension (J2000)                      &  08$^{h}\,$26$^{m}\,$51$\,\fs$489 \\
  Declination (J2000)                          &  $+$26$\degr\,$37$\arcmin \,$23$\,\farcs$706\\
  Epoch of frequency (modified Julian day)     &  54909.0                  \\
  Rotational frequency $\nu$ (Hz)              &  1.884439516337(1)         \\
  Rotational frequency derivative $\dot{\nu}$ (s$^{-2}$) &  -5.997(1)~$\times 10^{-15}$ \\
  Dispersion Measure $DM$~(cm$^{-3}$~pc)       &  19.464 \\
  \hline
\end{tabular}
\label{tab:0823par}
\end{table}

We find that there is no significant evidence for any periodicity in
the 153-d AFB timing residuals. However, this is not that surprising
considering the timescales of emission variation ($\sim$~hours) will
result in an observational bias against such detection; the ability to
resolve a discrepancy between observed and predicted TOAs will be
strongly dependent on the length of time a pulsar assumes a given
spin-down rate. Nevertheless, it is possible that the object
alternates between spin-down rates, consistent with the
magnetospheric-state changing scenario \citep{lhk+10}, but we are not
sensitive to the variations in the timing residuals. This idea will be
explored in greater detail in the following sections.

%\vspace{-5mm}
\subsection{Overview of the timing model}\label{sec:sims}
To determine whether the timing measurements of PSR~B0823$+$26 could
be consistent with a variable spin-down model, we developed a
simulation tool to reproduce the timing behaviour of intermittent
pulsars. This tool is based on a Monte-Carlo method, whereby we define
a parameter space of trial radio-on and -off spin-down rates
($\dot{\nu}_{\mathrm{on}}$ and $\dot{\nu}_{\mathrm{off}}$) to produce
simulated timing residuals that are compared with the observed. The
tool also uses an initial rotational frequency, $\nu_0$, and average
fitted rotational frequency derivative, $\dot{\nu}_{\mathrm{av}}$, to
simulate the timing behaviour of the object. These additional
parameters are determined from fitting the pulsar's rotational and
orbital parameters from the observed data using
{\fontfamily{cmr}\footnotesize\selectfont TEMPO2} (see
Table~\ref{tab:0823par}).

The emission activity of the source is represented by two analytic,
exponential functions that separately model the observed distributions
of emission phase durations (i.e. radio-on and -off). The pulsar is
modelled to sequentially switch between the radio-on and -off
phases. Therefore, the simulated emission phase durations are obtained
from the analytic functions by alternately and randomly sampling from
the possible ranges of values. This process is continued until the
emission activity of the total observed data duration ($\sim153$~d) is
fully covered for each simulation trial, thus producing pseudo-random
generated number distributions for both the radio-on and -off emission
phase durations. This method, in turn, provides us with a better
general description of the pulsar emission activity, due to the finite
number of data points we can sample from the parent
distributions. With the above in mind, the timing residuals of each
simulation trial represent one possible observed outcome, for a given
pair of spin-down rates, considering a pulsar which exhibits variable
emission activity with given distributions of switch durations.

As the spin-down rate of the model pulsar alternates between
consecutive emission phases, the rotational frequency is updated at
each integer step in pulse number:
\begin{equation}
\mathrm{TOA} = \frac{n}{\nu} + t_{\mathrm{ref}}\,.
\end{equation}
The reference time $t_{\mathrm{ref}}$ corresponds to the total time
elapsed at the last step and $n$ is the number of pulses in each
emission mode. The rotational frequency $\nu$ is updated using
\begin{equation}
\nu = \nu_{\mathrm{ref}} + (\dot{\nu}_{\mathrm{phase}} \times \Delta t)\,,
\end{equation}
where $\nu_{\mathrm{ref}}$ is the reference frequency at the last
step, $\dot{\nu}_{\mathrm{phase}}$ is the spin-derivative in the
corresponding emission phase and $\Delta t$ is the duration of the
emission mode. 

The simulated TOAs, obtained from these data, are those that would be
measured at the Solar System Barycentre (Barycentre reference frame)
and are inclusive of additive white Gaussian noise, to simulate
instrumental noise. For each simulated TOA, the noise signature is
calculated from a randomly selected error bar from the observed data;
a random number is generated from a Gaussian distribution with a
full-width at half-maximum that is twice the size of the error bar. To
track pulsar rotation in time, TOAs for radio-on and -off phases are
calculated. The radio-off phase TOAs are the theoretical TOAs which an
observer would measure if the pulsar was detectable.

We note here that, as $\nu$ is evolved over time, the initial value
for this quantity is equal to the rotational frequency at the start
point of each simulation trial $\nu_0$, which is obtained from the
average $\nu$ of the observed data. Therefore, for each combination of
$\dot{\nu}_{\mathrm{on}}$ and $\dot{\nu}_{\mathrm{off}}$, the
simulated value for $\nu_0$ will be different compared to the observed
due to the relative contributions of
$\dot{\nu}_{\mathrm{on\,,off}}$. This, in turn, results in a
systematic offset in the resulting timing residuals. To correct for
this effect, we fit an average $\nu$ to each set of simulated timing
residuals using {\fontfamily{cmr}\footnotesize\selectfont TEMPO2}. It
is important to note that we only consider the radio-on phase TOAs in
these fits, so as to accurately simulate the observations. Following
this procedure, we then use {\fontfamily{cmr}\footnotesize\selectfont
TEMPO2} to provide post-$\nu$-fit root-mean-square (hereafter, simply
referred to as RMS) values for each $\dot{\nu}_{\mathrm{on\,,off}}$
trial combination. The optimal combination of
$\dot{\nu}_{\mathrm{on}}$ and $\dot{\nu}_{\mathrm{off}}$ is then taken
as that which obtains an average RMS from these trial values which is
closest to the observed.

\subsection{Simulation analysis}
\begin{figure}
  \begin{center}
    % NB Trim dimensions are top,left,bottom,right
    \includegraphics[trim = 10mm 7mm 4mm 15mm,clip,height=8.3cm,width=5.5cm,angle=270]{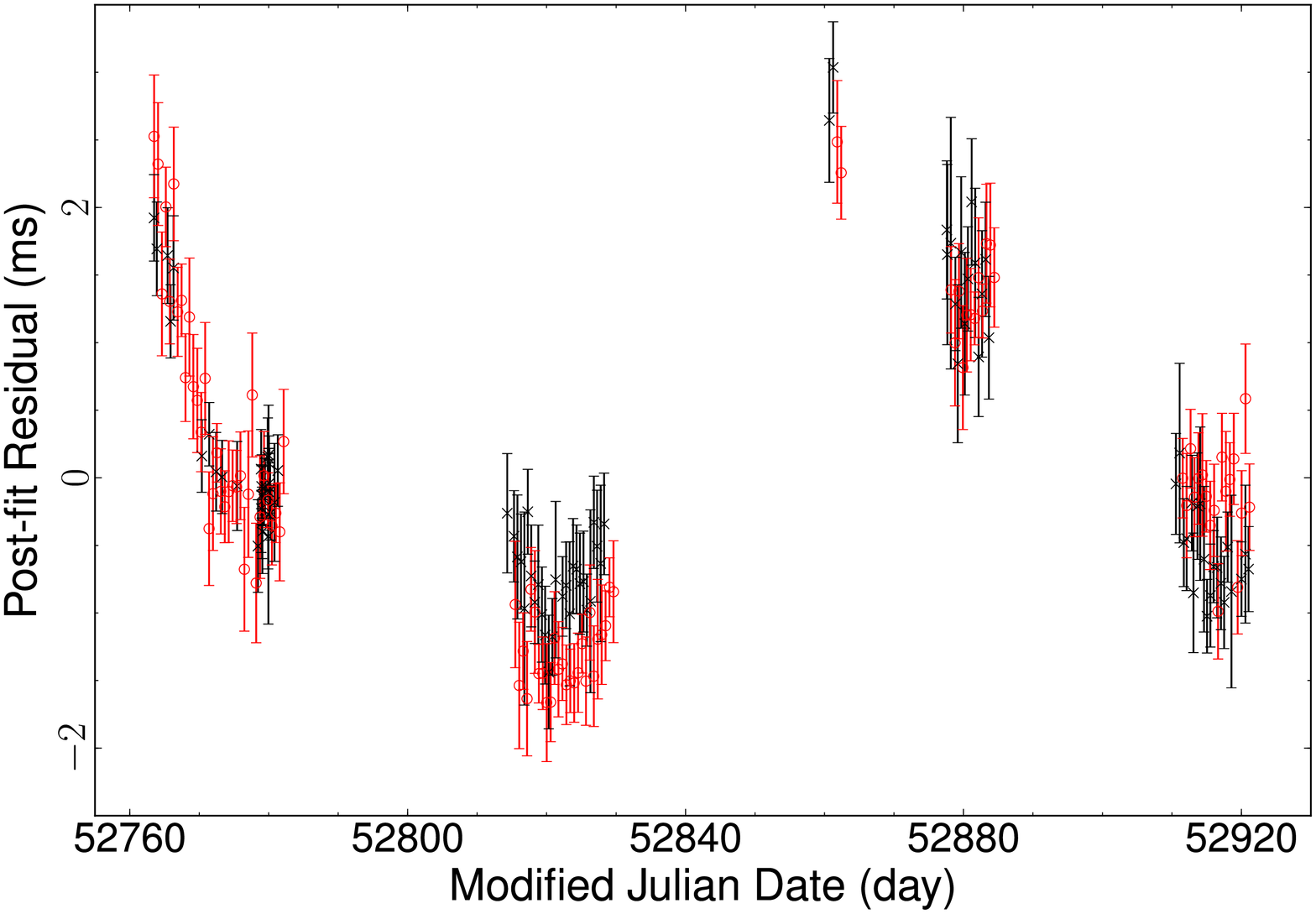}
  \end{center}
\vspace{-5pt}
\caption{Observed (\emph{crosses}) and simulated (\emph{open circles})
  timing residuals of PSR~B1931$+$24, using the same data as that in
  \protect\cite{klo+06} but with slightly different spin parameters. The
  results of the simulation are consistent with the timing behaviour
  of the source.}
\label{1931sim}
\end{figure}

To calibrate the simulation tool, and test its functionality, it was
applied to the prototype intermittent pulsar
PSR~B1931$+$24. \cite{klo+06} show that there is clear quasi-periodic,
cubic structure in the timing residuals of this pulsar and fit the
data to obtain
$\dot{\nu}_{\mathrm{on}}=-16.3~\pm~0.4~\times~10^{-15}$~s$^{-2}$ and
$\dot{\nu}_{\mathrm{off}}=-10.8~\pm~0.2~\times~10^{-15}$~s$^{-2}$. As
a consistency check, we analysed the same data-set using a number of
trial spin-down rates within the stated uncertainties. TOAs obtained
between 4~May~2003 and 9~October~2003 with the Lovell telescope were
analysed using {\fontfamily{cmr}\footnotesize\selectfont TEMPO2} to
determine $\nu_{0}$ and $\dot{\nu}_{\mathrm{av}}$, and to create an
accurate ephemeris, to perform the analysis. We find that the
simulation tool reproduces the timing behaviour of PSR~B1931$+$24
well, as shown in Fig.~\ref{1931sim}, for
$\dot{\nu}_{\mathrm{on}}=-16.1~\times~10^{-15}$~s$^{-2}$ and
$\dot{\nu}_{\mathrm{off}}=-10.8~\times~10^{-15}$~s$^{-2}$. We note
here that there are a couple of small offsets between the observed and
simulated residuals (at MJD$\sim52820$ and $\sim52918$), but emphasise
that these do not reflect on the validity of the simulation tool. We
stress that the example shown (Fig.~\ref{1931sim}) does not represent
a perfectly optimised result. It is also important to note that the
emission phase durations of PSR~B1931$+$24 are only known to an
accuracy of approximately $\pm1$~d. This, in turn, will naturally
result in a discrepancy between the observed and simulated data, if
the actual timescales of emission are different to those determined by
the observation sampling.

Following the above approach, we obtained $\nu_{0}$ and
$\dot{\nu}_{\mathrm{av}}$ for PSR~B0823$+$26. We determined the
constraints on the combinations of $\dot{\nu}_{\mathrm{on}}$ and
$\dot{\nu}_{\mathrm{off}}$ from the pulsar's ADC and average fitted
$\dot{\nu}$, which by definition is
\begin{equation}
\dot{\nu}_{\mathrm{av}} = (t_{\mathrm{on}} \times \dot{\nu}_{\mathrm{on}}) + (t_{\mathrm{off}} \times \dot{\nu}_{\mathrm{off}})\,,
\label{eq:nudot}
\end{equation}
where $t_{\mathrm{on}}$ and $t_{\mathrm{off}}$ are the fractional
times that the pulsar is in the radio-on and -off emission phases
respectively. Accordingly, the parameter space of spin-derivatives to
test was defined as
$-6.06~\leq~\dot{\nu}_{\mathrm{on}}~(10^{-15}$~s$^{-2})~\leq~-5.99$
and $-6.0~\leq~\dot{\nu}_{\mathrm{off}}~(10^{-15}$~s$^{-2})~\leq~-5.6$
(assuming a small variation in the spin-down rate between emission
phases). The spin-derivatives are inherently constrained by
\mbox{$\dot{\nu}_{\mathrm{av}}~\sim~-5.997~\times 10^{-15}$~s$^{-2}$},
so that neither $\dot{\nu}_{\mathrm{on}}$ or
$\dot{\nu}_{\mathrm{off}}$ can cross this boundary. For each
combination of input parameters, the RMS was calculated for 2000
iterations of the simulation tool using a resolution of
$2~\times~10^{-18}$~s$^{-2}$ in $\dot{\nu}_{\mathrm{on,\,off}}$. These
data were then averaged and compared with the RMS and measured
uncertainty from the observed data. The uncertainty was calculated
using {\fontfamily{cmr}\footnotesize\selectfont TEMPO2} to fit the
observed data across individual segments. Here, the average and
standard deviation of the sample population of observed RMS values
were computed, and then used to obtain the fractional error and
measured uncertainty in the RMS. Assuming a $3$-$\sigma$ error in the
observed RMS, we obtain a constraint of $\mathrm{RMS}~=~80~\pm10~\mu$s
on the simulated results.

\begin{figure}
  \begin{center}
    % NB Trim dimensions are top,left,bottom,right
    \includegraphics[trim = 30mm 1mm 12mm 0mm, clip, height=8.3cm,width=5.5cm,angle=270]{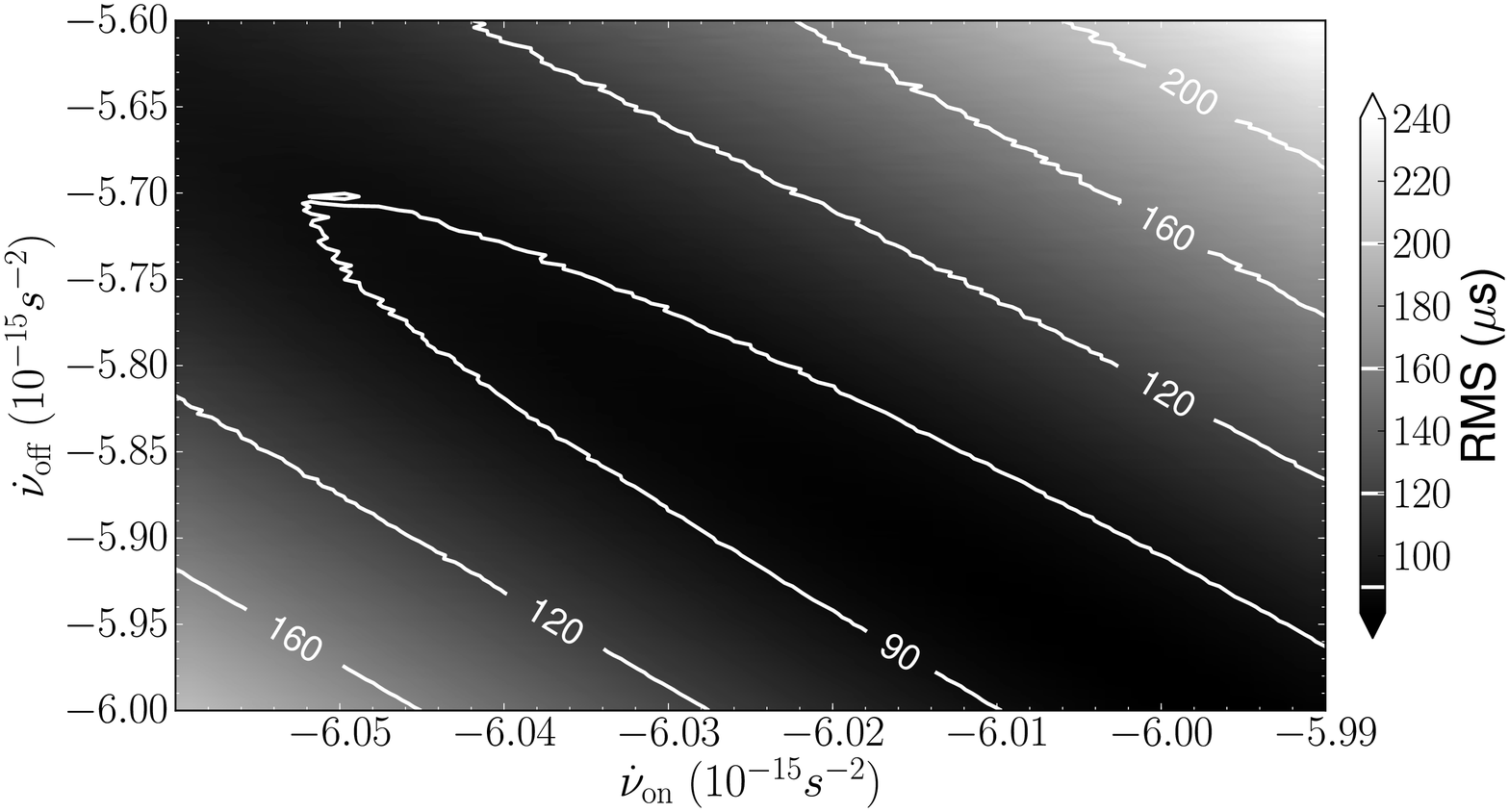}
  \end{center}
\vspace{-7pt}
\caption{Residual contour map obtained from simulations of the timing
  behaviour of PSR~B0823$+$26, for given combinations of rotational
  frequency derivatives. The plot shows the distribution of RMS values
  for the specified range of $\dot{\nu}_{\mathrm{on,\, off}}$, with
  the contours denoting RMS levels for the given parameters. The
  diagonal region in the middle of the plot shows the area of
  acceptable $\dot{\nu}$ combinations. The observed timing RMS is
  $80~\pm10~\mu$s.}
\label{0823cmap}
\end{figure}

The result of this analysis can be seen in Fig.~\ref{0823cmap}. The
parameter space of
$\dot{\nu}_{\mathrm{on}}~-~\dot{\nu}_{\mathrm{off}}$ where the RMS
converges on the observed value is seen to be distributed, within the
observational errors, towards the central diagonal portion of the
plot. This can be attributed to Eqn.~\ref{eq:nudot}, which shows that
$\dot{\nu}_{\mathrm{av}}$ of the simulated data, and hence the RMS by
logical progression, will converge with the observed for
$\dot{\nu}_{\mathrm{on}}\propto1/\dot{\nu}_{\mathrm{off}}$. Consequently,
we find that the maximum increase from $\dot{\nu}_{\mathrm{off}}$ to
$\dot{\nu}_{\mathrm{on}}$ ($\Delta \dot{\nu}_{\mathrm{max}}$) which
satisfies the model criteria is approximately $6\,\%$.

We also note that as the emission phase durations are sampled randomly
from the model distributions, there exists a variance in the computed
average RMS for each combination of input parameters. As such, we find
that the results of the simulation tool are less variable for
$\dot{\nu}$ values which are closer to the observed average; the
variance in the results from the simulated residuals has to be smaller
for the simulation to be consistent with the observations. Despite the
fact that the simulation tool can be applied to any intermittent
pulsar, it is ideally suited to modelling those with small $\Delta
\dot{\nu}$ as least-squares fitting methods, such as the method
implemented by \cite{klo+06}, will be more efficient for larger
$\Delta \dot{\nu}$. That is, the simulation tool can provide an upper
limit to $\Delta \dot{\nu}$, which is independent of curve-fitting
analyses which are unfeasible when analysing timing signatures with
negligible cubic structure.

\subsection{Switching timescale dependency}
Further to the analysis of the simulated data, the effects of altering
the emission phase durations and ADC were investigated. We were
interested in determining whether our results were subject to
systematic effects and how they were affected by modelling the
emission behaviour differently.

Firstly, we investigated the effect of over or under-estimating the
observed emission phase durations. We added systematic errors to the
observed durations, $t_{\mathrm{obs}}$, with a magnitude from
$0\,\%$~to~$100\,\%$ of $t_{\mathrm{obs}}$ in steps of $5\,\%$, with
the polarity of the error (for each emission mode) being determined by
a random number generator. As the percentage error was increased, the
average RMS over 2000 iterations of the simulation tool was found to
increase also. That is, as the variance in the emission phase
durations increases, so do the fluctuations in the timing residuals
due to the model pulsar assuming a spin-down rate for longer or
shorter than is observed. We find that the tolerance of the RMS to
these systematic errors increases towards smaller variations in the
spin-down rate. This is particularly true for the combinations of
spin-parameters which were comparable with the average spin-down rate;
the results of the simulation remained consistent with the observed
even for an error of $\pm~100\,\%$ in the emission phase durations. 

To confirm the validity of the maximum allowed change in spin-down
rate, $\Delta\dot{\nu}_{\mathrm{max}}\sim6\,\%$, we also performed a
simulation trial using the observed switching times without any
systematic errors. Here, we used the corresponding values for
$\dot{\nu}_{\mathrm{on\,,off}}$ which resulted in the maximum
spin-down rate variation. From this analysis, we obtain an RMS which
is greater than that observed. However, this result is only derived
from one trial and, as such, is not expected to be as consistent as
averaging over many simulated results. In addition, the value quoted
for $\Delta\dot{\nu}_{\mathrm{max}}$ inherently serves as an upper
limit. Therefore, it is possible that the pulsar may undergo more
modest variations in its spin-down rate; simulations incorporating
these spin-parameters produce results that are more consistent with
observations. We note, however, that systematic errors in the observed
emission durations will have a significant effect on the
$\Delta\dot{\nu}_{\mathrm{max}}$ limit obtained. As a result, we argue
that timing studies of intermittent pulsars will be more conclusive
for better sampled data-sets.

We also tested our assumption that the emission activity of
PSR~B0823$+$26 can be modelled by randomly sampling two distributions
of emission phase durations. Here, we compare the results of separate
trials which used the analytic and observed (both ordered and
randomised) sequences of emission phase durations. We find that the
average RMS values from these trials are all consistent. Therefore, we
conclude that the assumptions of our simulation tool are valid.

In addition, we examined the dependency of the results on the ADC
assumed. In this analysis, the sequences of radio-on and -off emission
phase durations were combined together and randomised, so that no
distinction was made between the different distributions. This, in
turn, resulted in an ADC which no longer coincided with the observed
value, that is $\sim50\,\%$. A number of $\dot{\nu}$ combinations were
used in this analysis, from $\Delta \dot{\nu}_{\mathrm{min}}$ to
$\Delta \dot{\nu}_{\mathrm{max}}$. However, no RMS values within the
observed $80\pm10~\mu$s could be obtained (except for
$\dot{\nu}_{\mathrm{on}}\sim\dot{\nu}_{\mathrm{off}}$). This indicates
that the form of the distribution which is used to model the emission
phase durations is fundamental to the simulation procedure.

To investigate this further, and to determine how unique the $\Delta
\nu_{\mathrm{max}}$ solution is, the analytic distributions of
emission phase durations were also directly altered. Here, simulations
were carried out where the parameters of the distributions were
modified to obtain two test cases, for $\sim5\,\%$ decrease and
increase in the ADC. These simulations returned
$\dot{\nu}_{\mathrm{on\,,off}}$ matches that corresponded best with
the ADC. That is, for an approximately $5\,\%$ decrease in the ADC,
the maximum change in the spin-down rate reduces, and vice versa. For
a lower percentage radio-on time, $\Delta \dot{\nu}$ is required to be
smaller to obtain $\dot{\nu}_{\mathrm{av}}$, and the opposite for a
higher percentage. As a result, we obtain limits on the maximum change
in spin-down rate \mbox{$4.5\,\% \lesssim
\Delta~\dot{\nu}_{\mathrm{max}} \lesssim 7.6\,\%$}, if we assume a
maximum error of $\sim5\,\%$ in the modelling of the random number
distributions. Consequently, we see that alteration to the ADC has a
significant effect on the results; if the initial parameters used are
wrong, the simulation tool will not converge on an optimum combination
of spin-down rates. Despite the incorporation of a $5\%$ uncertainty
in the ADC, it is apparent that the upper limits on the spin-down rate
variation are still small compared with that of PSR~B1931$+$24
($\sim50\,\%$; \citealt{klo+06}).

\section{Discussion}\label{sec:discuss}
\subsection{Nulling timescales in PSR~B0823$+$26}
While our data represent the most comprehensive study of the long-term
nulling behaviour of PSR~B0823$+$26, we suspect that the majority of
the inferred nulling timescales are influenced by data-windowing
effects. That is, we predict that the emission phase durations are
often overestimated due to insufficient time coverage. This is
strongly supported by the activity plots
(Fig.~\ref{0823onoff}~$-$~\ref{0823shortscale}) which, when compared,
clearly show that the pulsar exhibits a much shorter modulation
timescale ($\sim$~minutes to hours) than can be resolved in the
typical observing cadence ($\sim$~days). This effect also influences
the result of the WWZ analysis (Fig.~\ref{0823wwz}), due to the lack
of any significant spectral features at times of poor observation
cadence. Therefore, we cannot explicitly state the \emph{typical}
modulation timescale(s) intrinsic to PSR~B0823$+$26; it is likely that
the sparser sampling of the typical observations may have resulted in
missing numerous radio-on and -off states.

Although the majority of our data is subject to the data-windowing
effects mentioned above, the data from the continuous observing runs
is not. Therefore, we can use these data to place confident limits on
the short-term variability of PSR~B0823$+$26. In light of this, we
conclude that the source undergoes nulls over timescales of minutes to
hours (up to at least 5~h), but do not rule out the possibility of
longer ($\sim1$~d) null phases, which require further
investigation. Remarkably, in these observing runs, we also find
evidence for highly irregular short-term ($\lesssim1$~h)
modulation. That is, the object appears to exhibit periods of
spontaneous bursts of emission and nulls, or emission flickering,
before a constant emission phase is assumed. If this is typical for
neutron stars, it may be inferred that pulsars which exhibit
intermittency, e.g. normal nulling pulsars and rotating radio
transients (RRATs; \citealt{mll+06}), may undergo a radio emission
`ignition phase' before attaining the necessary criteria for stable
emission\footnote{In this context, RRATs may never attain the criteria
  for stable emission.}. However, this behaviour could just be
intrinsic to PSR~B0823$+$26 and, therefore, requires the study of
other sources to associate it to the pulsar population as a
whole. Evidently, this would be complemented by further investigation
into the typical nulling behaviour of PSR~B0823$+$26, which may
provide insight into this phenomenon and the mechanism(s) responsible
for emission modulation in pulsars.

Taking in mind that nulling pulsars exhibit null durations of
$1-10$~pulse periods up to approximately a year (PSR~J1841$-$0500;
\citealt{crc+12}), PSR~B0823$+$26 may be considered to be a bridge
between the different `types' of nulling pulsars. That is, the object
may be located inbetween conventional nulling pulsars and longer-term
intermittent pulsars on a `nulling continuum' scale \citep{kea10}. A
question to now ask then, is if these sources all show different
cessation timescales will all of them exhibit $\Delta\dot{\nu}$ and,
if so, how will it manifest? This will be investigated in greater
detail below.

\subsection{Evidence and implications of $\Delta\dot{\nu}$ in PSR~B0823$+$26}
For the first time, we have presented a simulation tool which can
model the timing behaviour of an intermittent pulsar and obtain an
upper limit on its spin-down rate variation. Through simulating the
rotational behaviour of PSR~B0823$+$26, we find $\Delta
\dot{\nu}_{\mathrm{max}}\sim6\,\%$ between emission phases, which
corresponds to
$\dot{\nu}_{\mathrm{on}}=-6.05~\times~10^{-15}$~s$^{-2}$ and
$\dot{\nu}_{\mathrm{off}}=-5.702~\times~10^{-15}$~s$^{-2}$. However,
we cannot discount a scenario where there is no variation in spin-down
rate between the different radio emission phases. Keeping this in
mind, we can still try to estimate the implied current flow in the
pulsar magnetosphere for both the radio-on and -off states using the
above values. We follow \cite{klo+06} and consider the simplest
possible emission model. That is, we assume that in the radio-off
state the pulsar spins down by a mechanism that does not involve
substantial particle ejection (e.g., via magnetic dipole radiation if
the pulsar is in vacuum). Whereas, in the radio-on state, we assume
that spin-down rate is enhanced by a torque from the current of an
additional plasma outflow. Modifications of this simple assumption are
possible (e.g.~considering the spin-down contribution of a
plasma-filled close-field line region, see \citealt{lst12a}), but for
our purposes the simplest model is sufficient. Hence, with the
assumption that the increase of the spin-down rate is purely due to
the torque of the charged plasma additionally existing in the radio-on
state, over the radio-off state vacuum spin-down, one derives an
estimate for the charge density of this wind component:
\begin{equation}
  \rho_{\mathrm{plasma}} \approx \frac{3I\,\Delta\dot{\nu}\,c^2}{1.26\times10^{21}
  R^6 \sqrt{\nu\dot{\nu}_{\mathrm{off}}}}\sim0.0024\,\textrm{C m}^{-3}\,,
\label{eq:rho_plasma}
\end{equation}
where $c$ is the speed of light in a vacuum, $R$ is the radius of the
pulsar (taken to be $10^6$~cm) and $I$ is the moment of inertia
(assumed to be $10^{38}$~kg~m$^2$). We note that this equation also
assumes that the object is an orthogonal rotator, i.e. the source has
a magnetic inclination angle of $90\degr$. While this assumption is
almost certainly not met for most sources, this seems to be indeed a
good approximation for PSR~B0823$+$26 ($\alpha\sim86\degr$;
\citealt{ew01})\footnote{In fact, we propose that deviation from a
  simple spin-down model could in principle be used to infer the
  magnetic inclination angle of a source in an independent
  fashion.}. Comparing the above estimate with the Goldreich-Julian
density, i.e. the screening density in a simple dipolar spin-down
model (e.g. \citealt{lk05})
\begin{equation}
  \rho_{\mathrm{GJ}} = \frac{B_{\mathrm{s}} \,\nu}{c}\sim0.0203\,\textrm{C m}^{-3}\text{,}
\label{eq:rho_gj}
\end{equation}
where $B_{\mathrm{s}}$ is the surface magnetic field strength of the
pulsar, we at least obtain a good indication that only a small amount
($\sim 12\,\%$) of the total available charge in the wind component is
contributing to the overall spin-down. This indicates that the open
field line region of the pulsar magnetosphere may never become
entirely depleted of charge during a radio-off phase
(c.f. \citealt{lst12a}) or, alternatively, that the spin-down of the
pulsar is dominated by the non-wind contribution (i.e. magnetic dipole
radiation).

For PSR~B1931$+$24 it is suggested that the radio emission cessation
results from re-configuration of the global magnetospheric charge
distribution, and that the spin-down rate variation is a signature of
this phenomenon \citep{klo+06,tim10,lhk+10,lst12a,lst12b}. Certainly,
in several other pulsars, we see evidence to support this theory
\citep{lhk+10}. However, it is not entirely clear what manner of
emission modulation will be accompanied by a variable spin-down rate;
in the study performed by \cite{lhk+10}, pulsars with similar changes
in spin-down rate were observed to undergo different magnitudes of
variation in their pulse shape. It is interesting, therefore, to find
that PSR~B0823$+$26 exhibits the same breakdown in radio emission
production (or detectability) as PSR~B1931$+$24, without the need for
a large change in the magnetospheric currents. This implies that the
mechanism which produces radio emission is highly sensitive to even
the smallest changes in the magnetosphere and that mode-changing and
nulling are closely related. From this, we infer that there could be a
significant number of pulsars that exhibit regulated changes in their
spin-down rate and emission, which contribute to `timing noise', and
that we are not yet aware of them due to small fractional changes in
$\dot{\nu}$ or short timescale variations.

\subsection{Detection limits on $\Delta\dot{\nu}$}
While a large number of pulsars are thought to exhibit spin-down rate
variation \citep{hlk10,lhk+10}, only 20 are known to exhibit any
discernible change in $\dot{\nu}$
(e.g. \citealt{klo+06,lhk+10,crc+12}). With this in mind, we sought to
determine our sensitivity to $\Delta\dot{\nu}$ in neutron stars and,
hence, whether we can expect to detect this effect in all
mode-changing and nulling pulsars.

\begin{figure}
  \begin{center}
    % NB Trim dimensions are top,left,bottom,right
    \includegraphics[trim = 18mm 16mm 6mm 24mm, clip, height=8.2cm,width=5.6cm,angle=270]{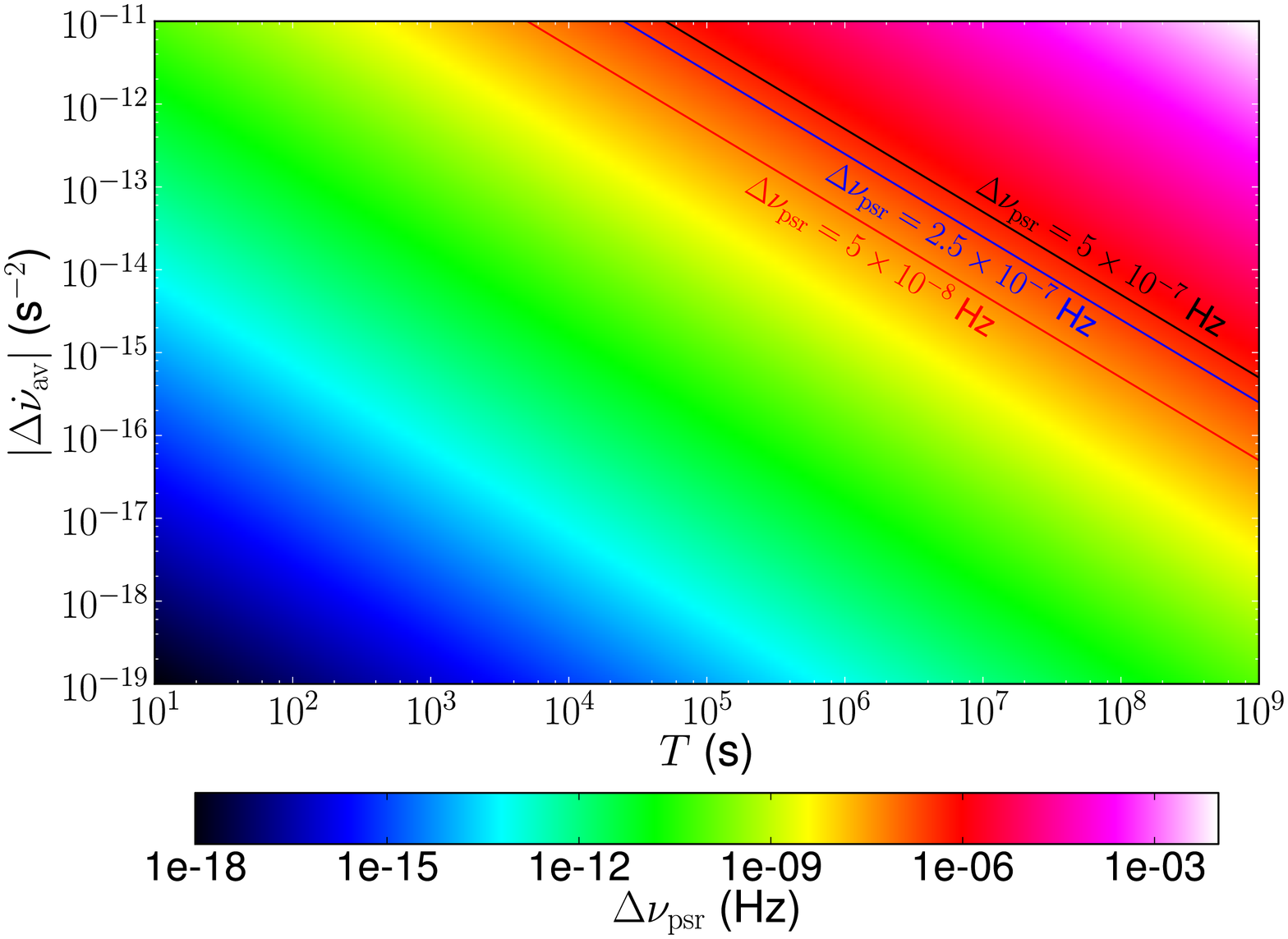}
  \end{center}
\vspace{-7pt}
\caption{The expected change in rotational frequency,
  $\Delta\nu_{\mathrm{psr}}$, due to an average spin-down rate
  deviation, $|\Delta\dot{\nu}_{\mathrm{av}}|$, that is assumed for a
  time $T$. Overlaid are lines denoting the possible detection
  requirements of a variable spin-down rate in PSR~B0823$+$26, that is
  (\emph{from left to right}) $\Delta\nu_{\mathrm{mod}}$,
  $5\,\Delta\nu_{\mathrm{mod}}$ and $10\,\Delta\nu_{\mathrm{mod}}$.}
\label{nudot_sens}
\end{figure}

We begin by considering a simple model of a dual-$\dot{\nu}$ nulling
pulsar, which alternately exhibits $\dot{\nu}_{\mathrm{on}}$ and
$\dot{\nu}_{\mathrm{off}}$ in its corresponding emission phases, for a
given length of time $\Delta t_i$. If the data is fitted with a
standard timing model with a single value of $\dot{\nu}$
($\dot{\nu}_{\mathrm{av}}=\dot{\nu}_{\mathrm{pred}}$; see, e.g.,
\citealt{mt77}), we expect the total difference between the observed
and predicted spin-down rates, in each emission phase, to be
$\Delta\dot{\nu}_i=\dot{\nu}_{\mathrm{obs}}-\dot{\nu}_{\mathrm{pred}}$.
Extending this model to $N$ emission phases, we would expect the total
discrepancy between the observed and predicted $\nu$ to be
\begin{align}
\Delta \nu_{\mathrm{psr}} &= \Delta \dot{\nu}_{\mathrm{on}}\,\Delta t_{\mathrm{on,\,1}} +  \Delta \dot{\nu}_{\mathrm{off}}\,\Delta t_{\mathrm{off,\,1}} + \Delta \dot{\nu}_{\mathrm{on}}\,\Delta t_{\mathrm{on,\,2}} + \cdots\nonumber\\
                          &= \sum_{i=1}^N \Delta\dot{\nu}_i\, \Delta t_i\,.
\end{align}
In light of this, it will be possible to detect the presence of a
variable $\dot{\nu}$ in the timing residuals of a pulsar when the
condition $\Delta \nu_{\mathrm{psr}}\gg\Delta \nu_{\mathrm{mod}}$ is
satisfied (where $\Delta \nu_{\mathrm{mod}}$ is the uncertainty in the
model $\nu$ over a time span $T$). With the above in mind, we computed
the expected $\Delta \nu_{\mathrm{psr}}$ over the course of three
subsequent emission phases (of total length $T$), assuming an average
deviation in spin-down rate
$|\Delta\dot{\nu}_{\mathrm{av}}|=|\langle\dot{\nu}_{\mathrm{obs}}-\dot{\nu}_{\mathrm{pred}}\rangle|$,
for a wide range of parameters (see Fig.~\ref{nudot_sens}). We note
that, while these data do not precisely model $\Delta
\nu_{\mathrm{psr}}$, they do offer useful estimates that can be used
to predict the detection limits on $\Delta\dot{\nu}$ in pulsars.

Assuming PSR~B0823$+$26 exhibits a maximum modulation timescale
$\Delta t\sim1$~d (hence $T\sim3$~d), and
$|\Delta\dot{\nu}_{\mathrm{av}}|\sim9.7\times 10^{-17}$~s$^{-2}$, we
would expect $\Delta\nu_{\mathrm{psr}}\sim2.5\times 10^{-11}$~Hz. This
is approximately 2000 times smaller than $\Delta\nu_{\mathrm{mod}}$
from our timing measurements ($\sim5\times 10^{-8}$~Hz), which can
clearly explain why we do not find direct evidence for multiple
$\dot{\nu}$ values. Comparing this with PSR~B1931$+$24, we find that
$\Delta\nu_{\mathrm{psr}}/\Delta\nu_{\mathrm{mod}}\sim3$, for
$|\Delta\dot{\nu}_{\mathrm{av}}|=8.8\times 10^{-16}$~s$^{-2}$ and
$\langle T \rangle\sim40$~d, does result in significant detection of a
variable $\dot{\nu}$. This indicates that our method of calculation
for $\Delta\nu_{\mathrm{psr}}$ is robust, and that it is not large
enough in PSR~B0823$+$26 to facilitate the detection of a variable
$\dot{\nu}$.

Moreover, considering that the maximum null duration resolved in
PSR~B0823$+$26 (during the continuous observing runs) is about 5~h, it
is likely that $\Delta\nu_{\mathrm{psr}}$ is overestimated by a factor
of $\sim5$. Therefore, it is not unsurprising that
$\Delta\nu_{\mathrm{psr}}$ is not significant enough to require the
incorporation of another $\dot{\nu}$ in the timing model. In light of
this result, it is extremely likely that $\Delta\dot{\nu}$ will not be
discernible in a significant proportion of nulling pulsars.

With the above in mind, it is interesting to determine the typical
properties of a pulsar which can facilitate the detection of
$\Delta\dot{\nu}$. If we assume a typical pulsar has
$\dot{\nu}_{\mathrm{av}}=-4.46\times 10^{-15}$~s$^{-2}$,
$|\Delta\dot{\nu}_{\mathrm{av}}|\sim4.5\times 10^{-16}$~s$^{-2}$
(i.e. $\sim10\,\%$ of $\dot{\nu}_{\mathrm{av}}$) and a detection
threshold $\Delta\nu_{\mathrm{psr}}/\Delta\nu_{\mathrm{mod}}\sim5$,
then we should expect to detect spin-down rate variation in pulsars
which exhibit $T\gtrsim26$~d and have a timing precision
$\Delta\nu_{\mathrm{mod}}\sim10^{-9}$~Hz or better. The majority of
mode-changing and nulling pulsars, however, do not exhibit such long
modulation timescales (e.g. \citealt{wmj07,kle+10,bb10,bbj+11,
kkl+11}). Accordingly, we argue that it will be extremely difficult to
determine whether $\dot{\nu}$ switching is a global phenomenon in
these sources, especially those with very short modulation timescales
(i.e. $\lesssim$~hours). Clearly, more observations of similar objects
are required before a consensus is reached on their common properties,
and how PSR~B0823$+$26 fits into their framework.

%\vspace{-5mm}
\section{Conclusions}\label{sec:conc}
We have found evidence to suggest that PSR~B0823$+$26 exhibits a broad
distribution of nulling timescales i.e. $\lesssim$~minutes up to
several hours or more. Although longer duration nulls ($\sim$~day)
have been observed in our data, we cannot rule out the presence of
short-term variations $\lesssim$~minutes in the radio-off phases which
we are insensitive to; radio emission flickering, which was observed
in the continuous observing runs as `pre-ignition' radio pulses, could
have occurred during times of low observation cadence and, hence, been
missed. As such, long ($\sim$~hours) single-pulse observations of this
source are required to confirm its overall nulling fraction, emission
intermittency timescales and, therefore, its periodicity; that is if
the source does not exhibit random emission fluctuations. These data
should also provide further information as to why nulling (a.k.a.
intermittent) pulsars might undergo emission flickering before they
assume a stable magnetospheric state.

A simulation tool was developed to reproduce the timing behaviour
observed in intermittent radio pulsars. This tool provides evidence
for an upper limit of approximately $6\,\%$ on the variation in the
spin-down rate of PSR~B0823$+$26. However, we cannot rule out a
scenario where the pulsar retains a constant, single
$\dot{\nu}$. Nevertheless, the low limit on $\Delta
\dot{\nu}_{\mathrm{max}}$, compared with that observed in
PSR~B1931$+$24 ($\sim50\,\%$; \citealt{klo+06}), suggests the
importance of small changes to the global charge distribution of a
pulsar magnetosphere, which might easily perturb the production or
detectability of radio emission. With this in mind, PSR~B0823$+$26
could provide a link between conventional nulling and more extreme
pulsar intermittency, due to the long timescale radio emission
modulation and small change in $\dot{\nu}$ between emission phases.

Despite the somewhat unique emission geometry of PSR~B0823$+$26 which
facilitates its detection as an inter-pulse pulsar, no morphological
clue has been provided to explain its intermittent behaviour. This is
primarily due to a dearth of studies which attempt to correlate pulsar
emission geometry with pulse intensity modulation, as well as the
ordinary location of the object in the $P-\dot{P}$ diagram which,
ultimately, make it extremely difficult to discern the general
properties (e.g. $P$, $\dot{P}$ and $\alpha$) of the source from the
bulk of the normal pulsar population. It is interesting to note,
however, that the object undergoes intermittent behaviour without the
need for large changes in spin-down rate, contrary to that predicted
by \cite{lst12a}
($\dot{\nu}_{\mathrm{on}}/\dot{\nu}_{\mathrm{off}}\sim120\,\%$ for
$\alpha\sim90\degr$). In this context, it may only be pulsars which
exhibit long modulation timescales ($\gtrsim$~days to weeks) that
experience such large changes in $\dot{\nu}$. Ultimately, further
(multi-frequency) study of PSR~B0823$+$26, and other objects like it,
should reveal significant information about the nature of normal
nulling pulsars, RRATs and longer-term intermittent pulsars. This, in
turn, should offer the possibility to determine how all these objects
are related (if they are), what their typical characteristics are and
what mechanism is governing their irregular behaviour.

%\vspace{-4mm}
\section{Acknowledgements}
We thank A.~A.~Zijlstra, A.~Shukurov, D.~J.~Champion and M.~D.~Gray
for useful discussions which have contributed to this work. We also
acknowledge C.~Jordan, and the several telescope operators at Jodrell
Bank, who have overseen the many hours of observations used in this
paper.

%\newpage
\bibliographystyle{mn2e} 
\bibliography{journals,psrrefs,njy_modrefs}
\end{document}